\documentclass[a4paper,11pt]{article}

\usepackage[english]{babel}

\usepackage[T1]{fontenc}

\usepackage{lmodern}
\usepackage{sectsty}
\usepackage{yfonts}
\allsectionsfont{\boldmath}

\usepackage{physics}
\usepackage{slashed}
\usepackage{makeidx}
\usepackage[vcentermath]{youngtab}
\usepackage{amssymb}
\usepackage{amsmath}
\usepackage{amsthm}
\usepackage{amsfonts}
\usepackage{mathtools}
\usepackage{braket}
\usepackage{enumerate}
\usepackage{comment}

\usepackage{setspace}
\usepackage[in]{fullpage}
\usepackage{hyperref}
\usepackage{array}
\usepackage{cancel}
\usepackage{wrapfig}
\usepackage[font=small,labelfont=bf]{caption}
\usepackage{pifont}

\hypersetup{
	colorlinks=true,
	linktoc=page,
	citecolor=blue,
	linkcolor=blue,
	urlcolor=blue} 

\usepackage{todonotes}

\usepackage{float}
\usepackage{marvosym}
\usepackage{empheq}
\usepackage{bbold}
\usepackage{fancyhdr}
\usepackage{fancyref}
\usepackage{varioref}

\usepackage{tikz}
\usetikzlibrary{shapes}
\usetikzlibrary{arrows}
\usetikzlibrary{positioning}
\usetikzlibrary{decorations.pathmorphing}
\usetikzlibrary{decorations.pathreplacing}
\usetikzlibrary{decorations.markings}
\tikzset{cross/.style={cross out, draw, 
         minimum size=2*(#1-\pgflinewidth), 
         inner sep=0pt, outer sep=0pt}}

\usepackage[nottoc,notlot,notlof]{tocbibind}
\usepackage[nosort]{cite} 

\newcommand{\agl}[2]{\langle#1 \, #2 \rangle}
\newcommand{\sqr}[2]{\lbrack #1 \, #2 \rbrack}

\newcommand{\C}{\mathbb{C}}
\newcommand{\so}{\text{SO}}
\newcommand{\slg}{\text{SL}}

\begin{document}


\renewcommand{\frefsecname}{Section}
\renewcommand*{\frefeqname}{eq.}


\begin{flushright}
	QMUL-PH-23-05\\
	SAGEX-22-33-E\\
\end{flushright}

\vspace{20pt} 

\begin{center}

	{\Large \bf \boldmath{\texttt{SpinorHelicity4D}: a Mathematica toolbox for}}  \\
	\vspace{0.3 cm} {\Large \bf the four-dimensional spinor-helicity formalism}

	\vspace{25pt}

	{\mbox {\sf  \!\!\!\!Manuel~Accettulli~Huber{\includegraphics[scale=0.05]{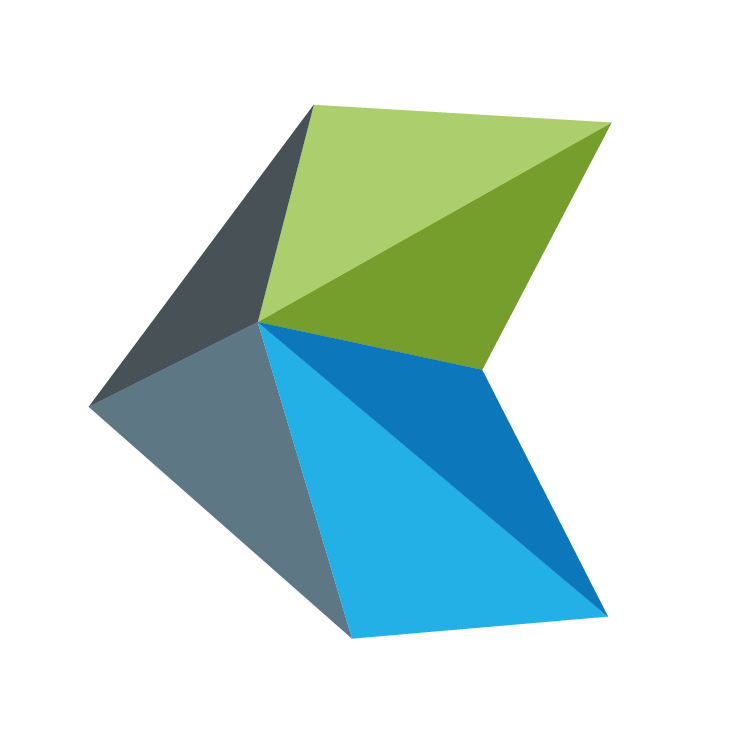}}
	}}
	\vspace{0.5cm}

	\begin{center}
		{\small \em
		Centre for Theoretical Physics\\
		Department of Physics and Astronomy\\
			Queen Mary University of London\\
			Mile End Road, London E1 4NS, United Kingdom
		}
	\end{center}


	\vspace{40pt}  

	{\bf Abstract}
\end{center}

\vspace{0.3cm}

\noindent
We present the \texttt{Mathematica} package \texttt{SpinorHelicity4D}, a dedicated suite for analytic and numeric calculations involving four-dimensional massless and massive spinor-helicity formalism. Analytic features of the package include for example: manipulation of contracted and uncontracted spinor quantities, automated application of Schouten identities for expression simplification, contractions of spinor products into chains, re-expression of chains in terms of Dirac traces and evaluation of such traces, derivatives of arbitrary functions of spinor quantities. Numeric features of the package include among others: generation of arbitrary $n$-point numerical complex kinematics, allowing for both massless and massive external states, fully numeric or parametric kinematics, and numeric generation on either $\mathbb{R}$ or $\mathbb{Q}$, the latter providing output immediately suitable for finite field applications. Furthermore, the package features userfriendly, intuitive but also highly customizable input options, thus providing an approachable tool for the casual user while still supporting more advanced applications for more adept and frequent users. All of the output is returned in the standard bracket notation, making it easily interpretable, but at the same it retains all of the analytic properties of the objects, allowing for copy-pasted and manipulated output to be provided as new input. This makes it ideal for front-end applications on a \texttt{Mathematica} notebook, while still allowing for deployment on a cloud server for more heavy calculations.

\vfill
\hrulefill
\newline
\vspace{-1cm}
${\includegraphics[scale=0.05]{Sagex.jpeg}}$~\!\!{\tt\footnotesize m.accettullihuber@qmul.ac.uk}

\setcounter{page}{0}
\thispagestyle{empty}
\newpage


\setcounter{tocdepth}{4}
\hrule height 0.75pt
\tableofcontents
\vspace{0.8cm}
\hrule height 0.75pt
\vspace{1cm}
\setcounter{tocdepth}{2}

\newpage

\section{Introduction}

The spinor-helicity formalism provides a parametrization of scattering amplitudes which, in many instances, is arguably the most convenient. A classic example is the Parke-Taylor \cite{Mangano:1990by} formula for MHV gluon-amplitudes, which in spinor notation reduces to a single term at any multiplicity $n$:
\begin{equation}\label{eq:ParkeTaylorEx}
	\mathcal{A}_n^{\rm tree}(1^+, \ldots, i^-, \ldots, j^-, \ldots, n^+) \> = \> i g^{n-2}\frac{\agl{i}{j}^4}{\agl{1}{2}\agl{2}{3}\cdots \agl{n}{1}} \> .
\end{equation}
This formalism is very flexible, and while originally it was mainly used for four-dimensional massless particles like in the Parke-Taylor case, it has since been adapted to a variety of different settings, including for example four-dimensional massive \cite{Kosower:2004yz, Arkani-Hamed:2017jhn} and six-dimensional massless particles \cite{Cheung:2009dc}. In this paper, we present a \texttt{Mathematica} package designed to deal with four-dimensional massless and massive particles, where in particular massive particles are represented through a pair of massless states.

In the design of the package, great effort has been put into keeping it accessible to the largest possible audience but without sacrificing the possibility of performing more advanced manipulations, which inevitably require larger amounts of technical knowledge. For the sake of userfriendliness four different ways of expression input are available, ranging from a very intuitive palette to the possibility of defining custom names for every object in the package. The latter functionality is aimed, among others, at providing a smooth transition from the use of other codes, be they private or other public packages like S@M \cite{Maitre:2007jq}, to the here presented routines. Output of the functions is always presented in the standard bracket notation. This was achieved through the use of \texttt{Mathematica}'s ``box'' functions, which provide the backbone of the front-end display of \texttt{Mathematica}'s own built-in objects. Consequently, the output representation does not affect the objects' analytic properties which are preserved at any time, allowing for output to be copied, pasted, edited and then independently re-run\footnote{This holds true both in \texttt{StandardForm} as well as \texttt{TraditionalForm}.}.

Input objects featured in the package include both the Lorentz-invariant spinor brackets and chains, as well as uncontracted covariant spinors and Levi-Civita tensors, and momentum and polarization vectors. Whenever uncontracted indices are present, standard contraction properties are automatically applied to form Lorentz-invariants, \textit{i.e.} $\lambda_1^{a}\lambda_{2\, a} \to \agl{1}{2}$ and $p^\mu q_\mu \to p \cdot q$. Intrinsic object properties are always accounted for, like for example antisimmetry of the spinor brackets or linearity of the scalar product. Some of these properties, including linearity among others, requires explicit declaration of momentum labels or of masslessness of states, so that no assumption is made internally by the code and the user retains full control over expressions. Whenever such a declaration is required, it is highlighted both in the attached example notebook as well as in this user guide by a light-gray cell background, and it is clearly stated in the function documentation. As one might expect, the main purpose of most of the analytic features in the package is to convert invariants among each other and apply properties so to simplify expressions. Some prime examples include \texttt{ChainSimplify}, which makes use of the Clifford algebra relation $\{\sigma^\mu, \bar{\sigma}^\nu\} = 2\eta^{\mu \nu}$ to simplify products of spinor brackets closed into continuos chains, and \texttt{SchoutenSimplify} which applies the Schouten identity $\agl{i}{j}\langle k | +\agl{j}{k}\langle i | + \agl{k}{i}\langle j |=0$. These routines are entirely automated and do not require the user to specify which particular replacement to perform in order to simplify a given expression.

Every object present in the package can also be numerically evaluated. This allows for quick comparisons even of very intricate analytic expressions, but it is especially useful when combined with functional reconstruction techniques (see \cite{Peraro:2016wsq}) or ansatz generation with consequent system solution (see \cite{AccettulliHuber:2021uoa}), allowing to recover analytic results from numerical evaluations. With this in mind, numeric kinematics is by default generated on $\mathbb{Q}$, ready to be mapped to finite fields. Numeric kinematics can be generated for a wide range of processes, including for example mixed massive-massless external states, or amplitudes on a unitarity cut where the momenta of some legs of one amplitude are related to momenta of another one.

The present user guide is structured into three parts, reflecting the three modules which compose the package. First, upon recalling some of the properties of the formalism and explaining how to get access to the package, we introduce the individual building blocks and their properties, all of which are defined in the module \texttt{SpinorBuildingBlocks} and their list can be accessed through the standard \texttt{?SpinorBuildingBlocks*}. Next, we go through the functions responsible for the manipulation of the analytic objects, which are defined in the module \texttt{SpinorHelicity4D}, and lastly we describe the numeric features provided by the module \texttt{SpinorNumerics}. As a final remark, it is important to stress that, despite most of the functions we present have been extensively tested and used by the author, bugs happen. We would be grateful for any bug-related flag which is raised, either through email or directly on github, but also gladly take any suggestion for improvement and welcome any further third-party development based on the functions here presented\footnote{We will happily expand the package through external code by the standard pull-request mechanics on git.}.




\section{Building blocks and expression input}

\subsection{A lightning review of the spinor-helicity formalism}\label{sec:theory}
In this first section we briefly review the four-dimensional spinor-helicity formalism, with the main
purpose of setting our notations and conventions. We limit our discussion to Weyl spinors only, which are our building blocks of choice.
For a more detailed review of the topic see for example \cite{Dixon:1996wi,Elvang:2013cua,Henn:2014yza,Brandhuber:2022qbk}, and for a broader amplitudes context see \cite{Travaglini:2022uwo}.

\subsubsection*{Definition of spinors}

The proper orthochronous Lorentz group $\so^+(1,3)$ admits a universal covering given by $\slg(2,\C)$,
consequently there is a one-to-one correspondence between projective representations of $\so^+(1,3)$
on the Hilbert space and the infinite-dimensional unitary representations of $\slg(2,\C)$. The states of the theory transform
under such unitary representations and induce the fields to transform under finite-dimensional (non-unitary) representation.
All the irreducible finite-dimensional representations of $\slg(2,\C)$ are labelled by a pair of
semi-integers $(m_L,m_R)$\footnote{Recall that
the algebra $\mathfrak{sl}(2,\C)$ is isomorphic to $\mathfrak{su}(2)_L \times \mathfrak{su}(2)_R$, and $(m_L,m_R)$ are related to the eigenvalues
of the Casimir operators $\mathbf{J}_{L/R}^2 = (J_{L/R}^1)^2+(J_{L/R}^2)^2+(J_{L/R}^3)^2$, with $J_{L/R}^i$ generators of the $\mathfrak{su}(2)_{L/R}$.
Elements of the two-dimensional vector space $\mathfrak{su}(2)_{L}$ will be denoted with $\lambda$ and elements of $\mathfrak{su}(2)_{R}$ with $\tilde{\lambda}$.},
and can be obtained from completely symmetrized tensor products of $2m_L$ copies of the fundamental and $2m_R$ copies of the anti-fundamental representations,
usually denoted as $\left(\frac{1}{2},0\right)$ and $\left(0,\frac{1}{2}\right)$.

The fundamental two-dimensional objects transforming in the $\left(\frac{1}{2},0\right)$ are denoted by $\lambda_{\alpha}$ and the associated Lorentz indices
are greek undotted letters, whereas $\tilde{\lambda}^{\dot{\alpha}}$ transform in the $\left(0,\frac{1}{2}\right)$ and the associated indices are dotted greek letters.
These objects will collectively be called (Weyl) \textit{spinors}.
The spinor indices can be raised and lowered by contraction with a two-dimensional Levi-Civita tensor as
\begin{equation}\label{eq:lamup}
	\lambda^\alpha=\epsilon^{\alpha \beta}\lambda_\beta \; ,\hspace{0.5cm} \tilde{\lambda}_{\dot{\alpha}}=\epsilon_{\dot{\alpha}\dot{\beta}}\tilde{\lambda}^{\dot{\beta}} \; ,
\end{equation}
where in our convention
\begin{equation}
\epsilon^{12}=-\epsilon_{12}=\epsilon^{\dot{1}\dot{2}}=-\epsilon_{\dot{1}\dot{2}}=1 \>,
\end{equation}
and
\begin{equation}
\epsilon_{\alpha \beta}\epsilon^{\beta \gamma}={\delta_\alpha}^\gamma \>.
\end{equation}
In order to avoid an often unnecessary cluttering of indices, one can introduce the shorthand bracket notation
\begin{equation}\label{eq:gpspinordef}
	\lambda_{i \, \alpha}=|\lambda_i\rangle \>,\hspace{0.5cm} \tilde{\lambda}_i^{\dot{\alpha}}=|\tilde{\lambda}_i]\> , \hspace{0.5cm} \hspace{0.5cm} \lambda_i^\alpha =\langle \lambda_i | \>, \hspace{0.5cm} \tilde{\lambda}_{i \, \dot{\alpha}} =[\tilde{\lambda}_i|\> \> .
	\end{equation}
It is then possible to build Lorentz invariant contractions as follows
\begin{equation}
	\def\arraystretch{1.5}
	\begin{array}{c}
		\agl{i}{j} := \agl{\lambda_i}{\lambda_j}:= \lambda_i^{\alpha} \lambda_{j \, \alpha}= -\agl{j}{i} \> , \\
		\sqr{i}{j} := \sqr{\tilde{\lambda}_i}{\tilde{\lambda}_j}:=\tilde{\lambda}_{i \, \dot{\alpha}}\tilde{\lambda}_j^{\dot{\alpha}}=-\sqr{j}{i} \> ,
	\end{array}
\end{equation}
which will be called \textit{angle} and \textit{square brackets} respectively.
Notice that given a triplet of spinors $\lambda_i^{\alpha}$, $\lambda_j^{\beta}$, $\lambda_k^{\gamma}$ any completely
antisymmetrized combination of them gives zero, thus upon contraction with a single Levi-Civita tensor one gets
the so called Schouten identity
\begin{equation}
	\agl{i}{j}\langle k | +\agl{j}{k}\langle i | + \agl{k}{i}\langle j |=0 \>,
\end{equation}
where the analogue for the $\tilde{\lambda}$ spinors clearly also holds.

\subsubsection*{Massless momenta}

Given a Lorentz four-vector $p^{\mu}$, it can be shown that the corresponding finite-dimensional representation of $\slg(2,\C)$
is $\left(\frac{1}{2},\frac{1}{2}\right)$. The map between the two representations can be explicitly realized through
\begin{equation}\label{eq:mommat}
	\def\arraystretch{1.5}
	\begin{array}{c}
		p_{\alpha \dot{\alpha}} := p_{\mu}\sigma^{\mu}_{\alpha \dot{\alpha}} \> , \\
		p^{\dot{\alpha} \alpha} := p_{\mu}\bar{\sigma}^{\mu \, \dot{\alpha} \alpha} \> ,
	\end{array}
\end{equation}
where given the Pauli matrices $\sigma^i$, we define $\sigma^{\mu}=\{ \mathbb{1}_2,\vec{\sigma} \}$ and $\bar{\sigma}^{\mu}=\{ \mathbb{1}_2,-\vec{\sigma} \}$,
which satisfy the Clifford algebra
\begin{equation}
	\{ \sigma^{\mu},\bar{\sigma}^{\nu}\}=2\eta^{\mu \nu} \>.
\end{equation}
If $p^{\mu}$ is the momentum associated to a particle of mass $m$, it is easy to see that
\begin{equation}
	m^2=p^2=det(p_{\alpha \dot{\alpha}})=\frac{1}{2}p^{\dot{\alpha}\alpha}p_{\alpha\dot{\alpha}} \>.
\end{equation}
If we consider now a \textit{massless particle}, the masslessness condition given by the vanishing of the determinant of $p_{\alpha\dot{\alpha}}$
is trivialized by setting
\begin{equation}\label{eq:palphaalphadot}
	p_{\alpha \dot{\alpha}}= \lambda_{\alpha}\tilde{\lambda}_{\dot{\alpha}} \>,
\end{equation}
due the antisymmetry of the angle and square brackets. Furthermore, notice that in momentum
space the massless Dirac equation translates into
\begin{equation}
	\begin{cases}
		p_{\alpha\dot{\alpha}}\lambda_p^{\alpha}=0 \\
		p_{\alpha\dot{\alpha}}\tilde{\lambda}_p^{\dot{\alpha}}=0 
	\end{cases}
\end{equation}
which is again automatically satisfied by writing the momentum as in \fref{eq:palphaalphadot}.

We can associate a vector $p^\mu$ to a momentum given in the spinor representation $p_{\alpha\dot{\alpha}}$ through the inverse map of \fref{eq:mommat}, given by
\begin{equation}\label{eq:momvec}
	p^\mu=\frac{1}{2}\langle p \, \sigma^\mu \, p]=\frac{1}{2} [p \, \bar{\sigma}^\mu \, p \rangle \>.
\end{equation}
When considering a momentum $k=-p$, it is easy to see from \fref{eq:palphaalphadot} that we can write the spinors $\lambda_k$ and $\tilde{\lambda}_k$ in terms of $\lambda_p$ and $\tilde{\lambda}_p$,
simply by defining
\begin{equation}\label{eq:negativemom}
	\lambda_{-p}^\alpha \equiv i \lambda_p^\alpha \>, \hspace{0.5cm} \tilde{\lambda}^{\dot{\alpha}}_{-p} \equiv i \tilde{\lambda}^{\dot{\alpha}}_p \>.
\end{equation}

The spinors $\lambda$ and $\tilde{\lambda}$ are in general complex-valued, and the reality condition on the momentum $p^\mu$ \fref{eq:momvec} translates into
\begin{equation}\label{eq:realitycond}
	\lambda=\tilde{\lambda}^*
\end{equation}
up to an arbitrary phase which we set to 1. Usually one is rather lenient towards this condition,
since it turns out that, both in some analytic as well as numeric settings (see \fref{sec:numerics}), it is very convenient to allow
for complex momenta\footnote{Or alternatively to consider a different space-time signature \cite{Witten:2003nn}, which similarly invalidates \fref{eq:realitycond}.}.
Notice that there is no unique way of associating spinors to a given momentum $p^\mu$,
since the rescaling
\begin{equation}\label{eq:littlegroup}
	\lambda \mapsto t \, \lambda \> , \hspace{0.5cm} \tilde{\lambda} \mapsto \frac{1}{t}\tilde{\lambda}
\end{equation}
clearly leaves the momentum invariant. Here $t \in \C$ in general whereas it is just a complex phase if \fref{eq:realitycond} applies.
\Fref{eq:littlegroup} is called little group scaling, and it implements at the level of spinors those Lorentz
transformations which preserve the given momentum.

When discussing processes involving particles of spin one, polarization vectors are usually required.
In our conventions these can be written as
\begin{equation}
    \varepsilon_+^{\dot{\alpha}\alpha}(p,r)=\sqrt{2} \, \frac{\tilde{\lambda}_{p}^{\dot{\alpha}}\lambda_r^\alpha}{\agl{r}{p}}=\sqrt{2} \, \frac{|p]\langle r|}{\agl{r}{p}} \>, \hspace{0.5cm}
    \varepsilon_-^{\dot{\alpha}\alpha}(p,r)=\sqrt{2} \, \frac{\tilde{\lambda}_{r}^{\dot{\alpha}}\lambda_p^\alpha}{\sqr{p}{r}}=\sqrt{2} \, \frac{|r]\langle p|}{\sqr{p}{r}} \>,
\end{equation}
or equivalently as vectors
\begin{equation}\label{eq:polarizations}
    \varepsilon_+^\mu (p,r)=\frac{1}{\sqrt{2}} \frac{\langle r \, \sigma^\mu p ]}{\agl{r}{p}} \> , \hspace{0.5cm}
    \varepsilon_-^\mu (p,r)=\frac{1}{\sqrt{2}} \frac{\langle p \, \sigma^\mu r ]}{\sqr{p}{r}} \> .
\end{equation}
It can be checked that these satisfy all the properties required for polarization vectors (in the Lorentz gauge):
\begin{equation}
    \def\setlength{2}
    \def\arraystretch{1.5}
    \begin{array}{cc}
        \varepsilon_+(p,r)^*=\varepsilon_-(p,r) \>, & p_\mu \, \varepsilon_\pm^\mu(p,r)=0 \>, \\
        |\varepsilon_\pm(p,r)|^2=-1 \>, & \varepsilon_+(p,r) \cdot \varepsilon_-(p,r)^*=0 \>.
    \end{array}
\end{equation}

Given two massless momenta $p_i$ and $p_j$, the associated Mandelstam invariant is defined as
\begin{equation}
	s_{ij}:=(p_i+p_j)^2=2p_i \cdot p_j=p_{i}^{\dot{\alpha}\alpha}p_{j\, \alpha\dot{\alpha}}=\agl{i}{j}\sqr{j}{i} \>,
\end{equation}
where, using $|p \rangle [p|=p$ for massless momenta, we can rewrite
\begin{equation}
	\agl{i}{j}\sqr{j}{i}  = \langle i \, j \, i] \>. 
\end{equation}
We will refer to structures such as this as chains. Other examples of chains include
\begin{equation}
	\begin{split}
		\langle q \, p_1 \ldots p_{2n} \, k \rangle &=-\langle k \, p_{2n} \ldots p_{1} \, q \rangle \\
		[q \, p_1 \ldots p_{2n} \, k ] &=-[k \, p_{2n} \ldots p_{1} \, q ] \\
		[q \, p_1 \ldots p_{2n+1} \, k \rangle &= \langle k \, p_{2n+1} \ldots p_1 \, q ]
	\end{split}
\end{equation}
where the $p_i$ are not necessarily massless.

\subsubsection*{Massive momenta}

Considering massive particles, while \fref{eq:mommat} still applies, \fref{eq:palphaalphadot} does not hold anymore.
It is nonetheless still possible to write the momentum directly in terms of the spinors $\lambda$ and $\tilde{\lambda}$,
by simply considering it as a linear combination of two massless momenta \cite{Kosower:2004yz}
\begin{equation}\label{eq:mommatmassive}
	P^\mu:=q^\mu+\frac{m^2}{2q\cdot k}k^\mu \hspace{0.5cm} \rightarrow \hspace{0.5cm} P_{\alpha\dot{\alpha}}=\lambda_\alpha\tilde{\lambda}_{\dot{\alpha}}+\frac{m^2}{\agl{k}{q}\sqr{q}{k}}\mu_\alpha\tilde{\mu}_{\dot{\alpha}} \>,
\end{equation}
with $q^2,k^2=0$ and $q_{\alpha\dot{\alpha}}=\lambda_\alpha\tilde{\lambda}_{\dot{\alpha}}$ and $k_{\alpha\dot{\alpha}}=\mu_\alpha\tilde{\mu}_{\dot{\alpha}}$.
Notice that counting the number of (real) degrees of freedom of the spinors in \fref{eq:mommatmassive} one finds three too many.
These are spurious degrees of freedom corresponding to the little-group transformations of the massive momentum, which in
four dimensions is implemented by an $\so(3)$ subgroup of the Lorentz group.
Just as the massless spinors defined up to \eqref{eq:littlegroup} are equivalent, one can
use these spurious degrees of freedom to set the spinors associated to $k$ to arbitrary values.
We will refer to $\mu$ and $\tilde{\mu}$ as reference spinors.

From \fref{eq:mommatmassive} one can define a pair of spinors associated to the massive momentum $P$
\begin{equation}
    |P\rangle = |q \rangle +\frac{m}{\sqr{q}{k}}|k] \>, \hspace{0.5cm} |P]=|q]+\frac{m}{\agl{q}{k}}|k\rangle \>,
\end{equation}
which have very different properties compared to the massless spinors, for example if $P$ and $K$ are massive then in general $\langle P K]\neq 0$.
The advantage of the decomposition of \fref{eq:mommatmassive} is that it allows to recycle much of the technology introduced for massless particles.
The price to pay however is that certain symmetries of the amplitude are obscured, in particular
the covariance of the amplitude under little group transformations of the massive momenta.
To make such property manifest it is more convenient to introduce a new set of spinors $\lambda^I_\alpha$, $\tilde{\lambda}^I_{\dot{\alpha}}$, carrying an additional SU(2)
index\footnote{While for massless spinors the little group is just an U(1) phase, for massive particles in four dimensions is given by SU(2).},
accounting explicitly for little group transformations \cite{Arkani-Hamed:2017jhn}. We then have
\begin{equation}
    P_{\alpha\dot{\alpha}}=\lambda^I_\alpha\tilde{\lambda}_{\dot{\alpha}I}=\epsilon_{IJ}\lambda^I_\alpha\tilde{\lambda}^J_{\dot{\alpha}} \>,
\end{equation} 
and the massive equivalent relation of \eqref{eq:littlegroup} reads $\lambda^I_\alpha \to U^I_J \lambda^J_\alpha$ for $U^I_J \in$ SU(2).
It can be shown that the previously introduced spinors $|P\rangle$ and $|P]$ simply correspond to a specific choice of
the $\lambda^I_\alpha$ and $\tilde{\lambda}^I_{\dot{\alpha}}$, where the little group index has been fixed. We live the implementation of this
covariant formalism for future work.



\subsection{Installation of the package}

The package, which is available on the author's github\footnote{Link to package repository: \href{https://github.com/accettullihuber/SpinorHelicity4D}{https://github.com/accettullihuber/SpinorHelicity4D}.}, consists of one main script and two dependencies:
\begin{itemize}
	\item \texttt{SpinorHelicity4D} is the main script and is the only one the user needs to call. This script contains all the functions dealing with analytic manipulations of the spinor expressions.
	\item \texttt{SpinorBuildingBlocks} is the dependency where the spinor and vector building blocks are defined, along with some related functions. No user interaction with this module is required.
	\item \texttt{SpinorNumerics} is the dependency where everything related to generation of numeric kinematics is defined, again no user interaction with this module is required.
\end{itemize}
In order to use the package, all three modules need to be installed, and this can be achieved in different ways, we provide here three examples:
\begin{enumerate}
	\item (recommended) clone the git repository to your local machine and add the folder path to \texttt{Mathematica}'s \texttt{init.m} file.
	\begin{enumerate}
		\item Instructions on how to clone a repository can be found on the \href{https://docs.github.com/en/repositories/creating-and-managing-repositories/cloning-a-repository}{official github documentation}. In order to manage your repositories easily we suggest to use a git GUI like for example \href{https://desktop.github.com/}{``Github Desktop''}. Cloning the \texttt{SpinorHelicity4D} repository instead of simply downloading its content, will allow you to stay up to date with future releases or possible bug fixes. Once the repository has been cloned it will be available as a folder on your local machine at a path of your choosing, let's call it \verb|M:\your\path\SpinorHelicity4D|.
		\item In order to access the package within any \verb|Mathematica| session without the need of specifying the package location every time, we recommend to add the line of code \verb|AppendTo[$Path,"M:\your\path\SpinorHelicity4D"]| to the \texttt{init.m} file in the \texttt{Kernel} folder, whose location can be obtained by typing \verb|$UserBaseDirectory| or \verb|$BaseDirectory| in a \verb|Mathematica| notebook.
	\end{enumerate}

	\item Download the repository's content and add the folder path to \texttt{Mathematica}'s \texttt{init.m} file:
	\begin{enumerate}
		\item Download the repository and save the three \verb|.wl| file components to a folder at a location of your choosing, let's call it \verb|M:\your\path\SpinorHelicity4D|. Opting for this method will require you to download the pacakge from scratch in order to get access to updates and new releases.
		\item Edit the \texttt{init.m} file as described in 1.b above.
	\end{enumerate}
	
	\item Download the repository's content and copy the three \verb|.wl| files into the \texttt{Applications} folder of your \texttt{Mathematica} installation, whose location is again found at \verb|$UserBaseDirectory| and \verb|$BaseDirectory|. The advantage of this approach is that no editing of the \verb|$Path| variable is required, since by default the \texttt{Applications} folder is where \texttt{Mathematica} looks for packages. The disadvantage is that in order to get access to updates or future releases will require you to download the package from scratch and repeat the procedure.

\end{enumerate}

Once any of the above procedures was carried out successfully, the package can be accessed by simply typing \verb|<<SpinorHelicity4D|, as shown below.
\begin{figure}[H]
	\begin{center}
		\makebox[\textwidth]{
			\fbox{\includegraphics[page=1,trim={1.5cm 21.5cm 4cm 2.5cm},clip]{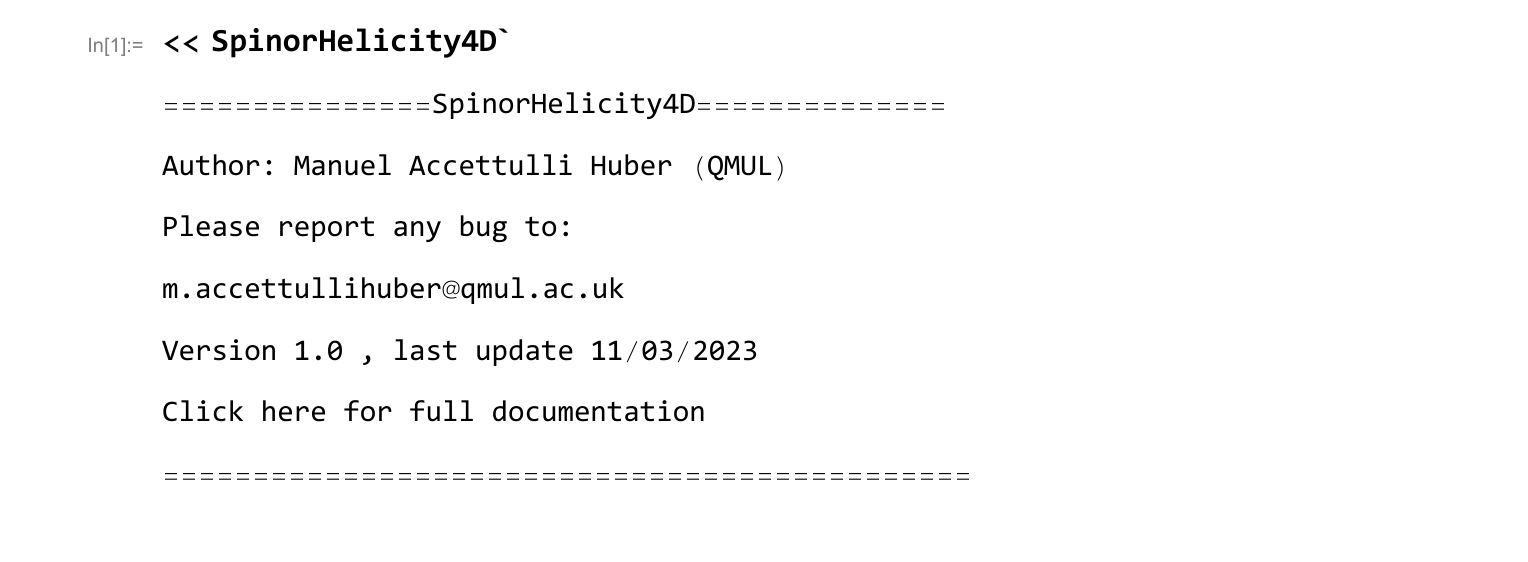}}}
	\end{center}
\end{figure}
In order to access a list of the functions in the package, type \verb|?PackageName*| replacing PackageName with the name of the appropriate module, depending on whether you look for analytic, numeric or building block functions.

\subsection{Expression input}

There are four different but completely equivalent ways for inputting expressions.
\begin{itemize}
	\item Through direct input of the function name.
	\begin{figure}[H]
		\begin{center}
			\makebox[\textwidth]{
				\fbox{\includegraphics[page=1,trim={1.5cm 21.5cm 4cm 5cm},clip]{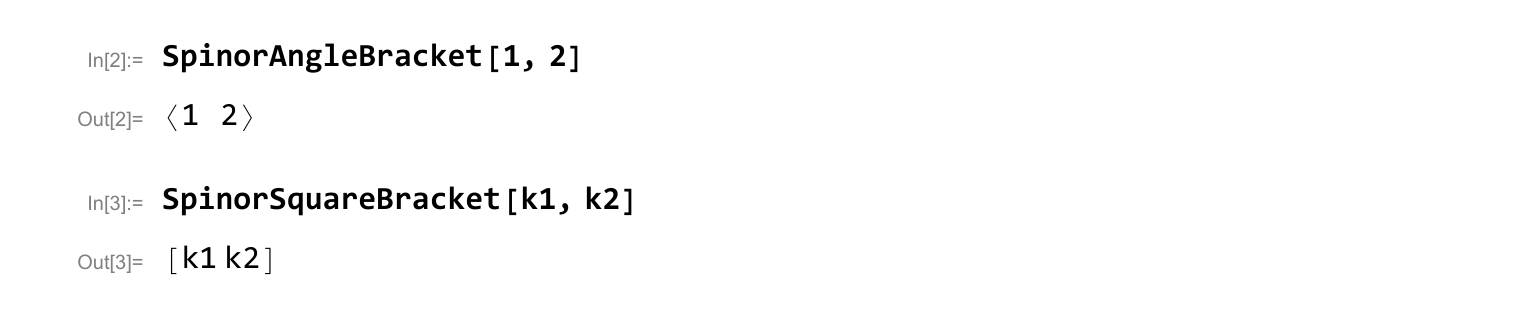}}}
		\end{center}
	\end{figure}
		\item Through an input palette, consisting of a small external notebook with a collection of control buttons. This palette will appear
			automatically when the package is loaded, but it can also be accessed by typing \texttt{SpinorPalette[]}. Clicking on a given button 
			will produce the corresponding object in the notebook, one can then navigate through the placeholders with the TAB keyboard command replacing them with the desired momenta/indices.
	\item Through keyboard shortcuts associated to every building block, consisting of a short string which, when escaped, produces the desired input. A list of all the available 
	shortcuts can be accessed by typing \texttt{\$Shortcuts}.
	\item Through customized definitions of the functions' names with \texttt{AssignNames}.
	For example, an accustomed S@M \cite{Maitre:2007jq} user could define the angle and square brackets as \texttt{Spaa} and \texttt{Spbb} respectively.
	This is achieved as shown below.
	\begin{figure}[H]
		\begin{center}
			\makebox[\textwidth]{
				\fbox{\includegraphics[page=1,trim={1.5cm 3cm 4cm 20cm},clip]{Example_building_blocks}}}
		\end{center}
	\end{figure}
	The difference between using \texttt{AssignNames} and directly writing the definitions in the notebook is that the former creates a file
	\footnote{This file is located in the same folder as the package.}, where the definitions are stored, and then loaded along with the package every time this is called. The definitions become thus permanent. 
	\begin{figure}[H]
		\begin{center}
			\makebox[\textwidth]{
				\fbox{\includegraphics[page=1,trim={1.5cm 23.5cm 4cm 2.5cm},clip]{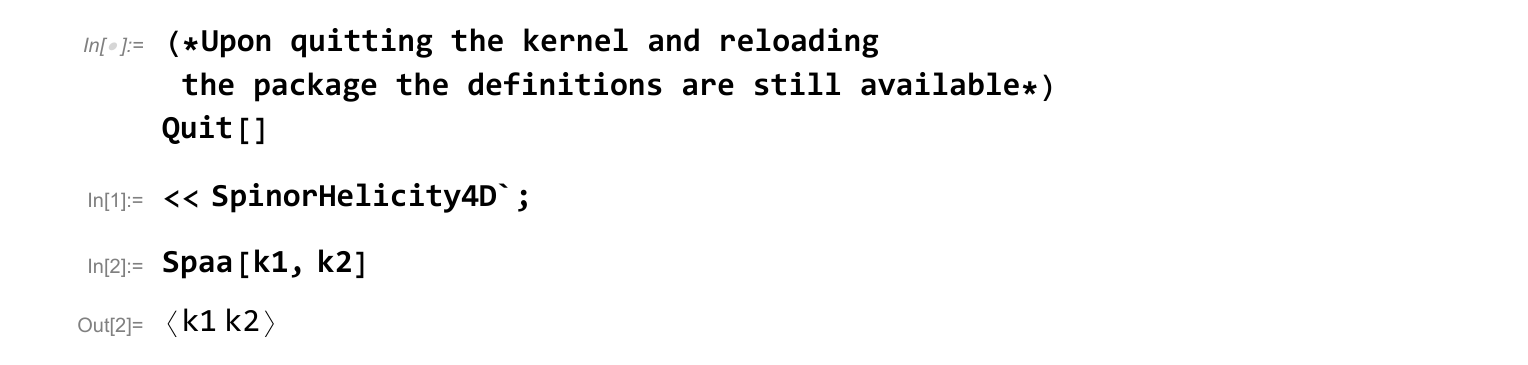}}}
		\end{center}
	\end{figure}
		The definitions can be cleared using \texttt{ClearNames}, which can be called either with one or more arguments to clear the desired functions' definitions,
	or without arguments to clear them all.
	\begin{figure}[H]
		\begin{center}
			\makebox[\textwidth]{
				\fbox{\includegraphics[page=1,trim={1.5cm 16.8cm 4cm 7.8cm},clip]{Example_building_blocks2}}}
		\end{center}
	\end{figure}
	\end{itemize}

\subsection{Declaration of momentum labels}

Many of the functions defined in the package posses definite properties with respect to their arguments.
For example the scalar product, which we call \texttt{mp}, is clearly linear with respect to four-vectors, thus given three momenta $p,q,k$ and a constant $c$
one expects
\begin{equation}
	p\cdot (q+c\, k)=p\cdot q +c \, p\cdot k  \>. 
\end{equation}
In order for this and other similar properties to be applied correctly in Mathematica, it is necessary to first declare which symbols have to be treated as particle/momentum labels.
This is achieved through \texttt{DeclareMom}:
\begin{figure}[H]
	\begin{center}
		\makebox[\textwidth]{
			\fbox{\includegraphics[page=1,trim={1.5cm 15.3cm 4cm 8.5cm},clip]{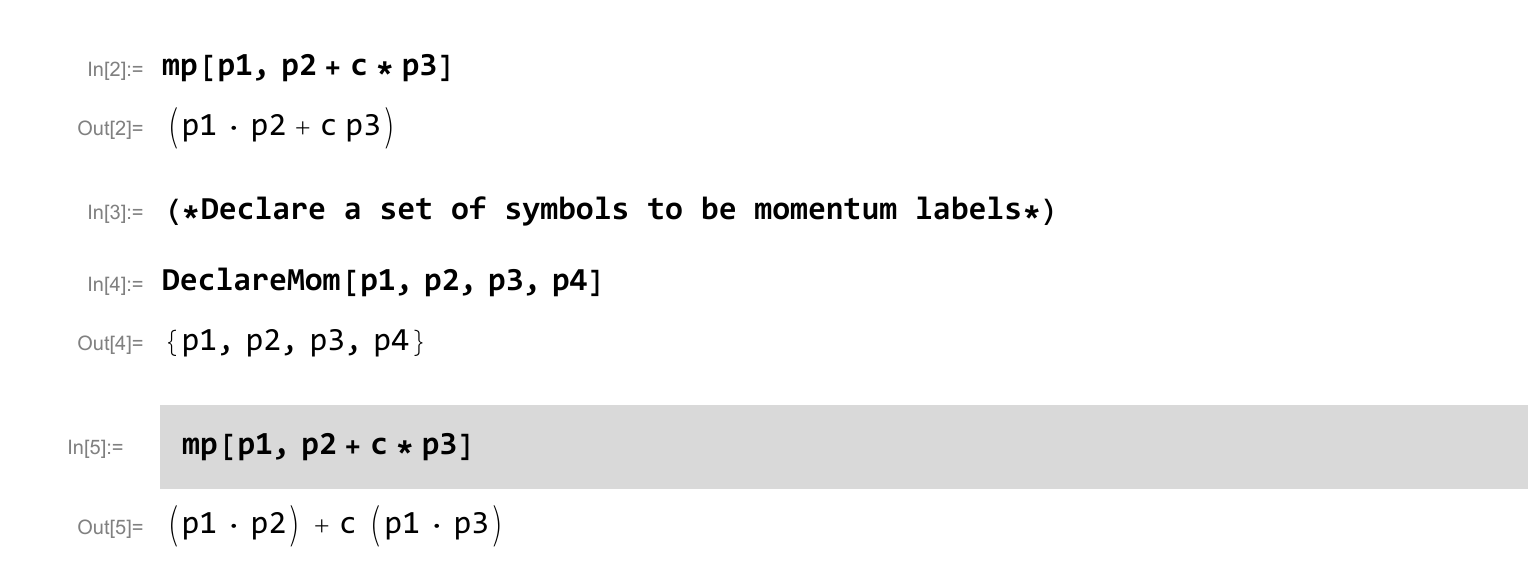}}}
	\end{center}
\end{figure}
From here on, whenever momentum declaration is required to obtain the displayed output, we will
use the light gray background for the associated input. It is worth pointing out that declaration of momentum labels is only necessary if access to the full properties of functions and objects
is needed, thus the basic usage of the package retains an intuitive nature. Notice also that it is not possible to declare numbers as momentum labels.
The list of so far declared labels can be accessed by calling \texttt{DeclareMom} without any arguments, and labels can be removed from this list with \texttt{UndeclareMom}.
\begin{figure}[H]
	\begin{center}
		\makebox[\textwidth]{
			\fbox{\includegraphics[page=1,trim={1.5cm 9cm 4cm 14.2cm},clip]{Example_building_blocks3}}}
	\end{center}
\end{figure}

\subsection{Declaration of massless momenta}
In general, since the package is designed in order to be able to handle both massless and massive
momenta, no assumption is made on momentum labels. Thus, in order to access the full range of simplifications
occurring due to a momentum being massless, one has to declare it as such through DeclareMassless. To remove labels from 
the list of massless momenta use \texttt{UndeclareMassless}, and to clear them all \texttt{UndeclareMassless} without any arguments, just as in the case of \texttt{UndeclareMom}.
\begin{figure}[H]
	\begin{center}
		\makebox[\textwidth]{
			\fbox{\includegraphics[page=2,trim={1.5cm 12cm 4cm 9.3cm},clip]{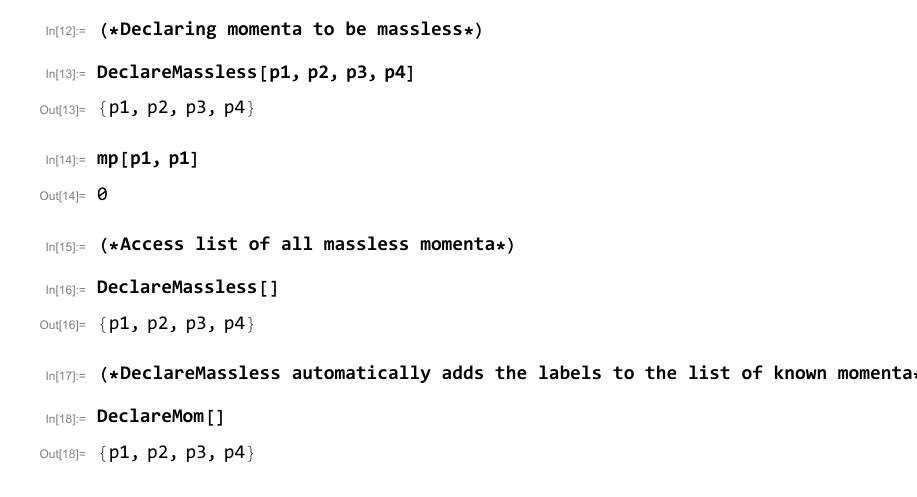}}}
	\end{center}
\end{figure}

\subsection{Angle and square brackets}

Angle and square brackets are entered through the associated functions \texttt{SpinorAngleBracket} and \texttt{SpinorSquareBracket} respectively,
or equivalently using the palette or the shortcuts.
\begin{figure}[H]
	\begin{center}
		\makebox[\textwidth]{
			\fbox{\includegraphics[page=1,trim={1.5cm 21.5cm 4cm 3.5cm},clip]{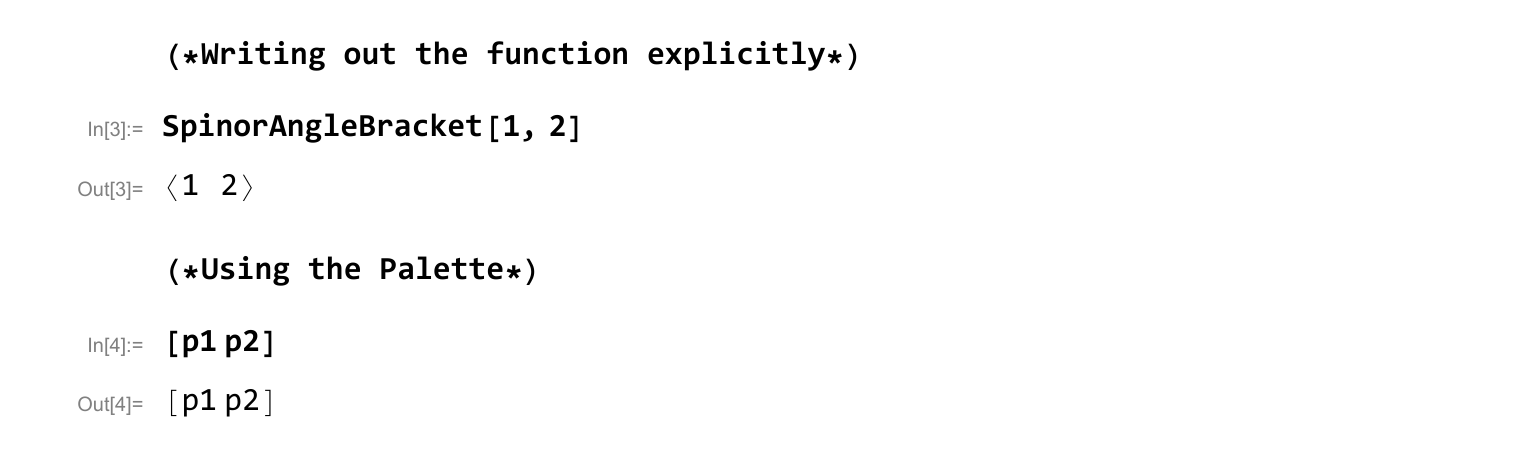}}}
	\end{center}
\end{figure}
Here it is important to stress that even when displayed as $\agl{1}{2}$, this object is still recognised as the function \texttt{SpinorAngleBracket[1,2]},
and thus it can for example be safely copy-pasted or otherwise manipulated. Once momentum labels are defined,
the convention on negative momenta \fref{eq:negativemom} is automatically applied.
\begin{figure}[H]
	\begin{center}
		\makebox[\textwidth]{
			\fbox{\includegraphics[page=1,trim={1.5cm 14cm 4cm 12.8cm},clip]{Example_building_blocks4}}}
	\end{center}
\end{figure}

The arguments of the spinor brackets are ordered according to Mathematica's canonical ordering,
taking into account antisymmetry.
\begin{figure}[H]
	\begin{center}
		\makebox[\textwidth]{
			\fbox{\includegraphics[page=1,trim={1.5cm 9cm 4cm 17cm},clip]{Example_building_blocks4}}}
	\end{center}
\end{figure}

Spinors are generally not linear with respect to their momentum argument, in other words
\begin{equation}\label{eq:pseudolinearity}
	\lambda^\alpha_{p+q} \neq \lambda^\alpha_{p} + \lambda^\alpha_{q} \>,
\end{equation} 
however when typing into the computer it is usually very convenient to apply the following definition:
\begin{equation}\label{eq:linearitydef}
	\agl{p_1}{(p_2+c\, p_3)} \equiv \agl{p_1}{p_2}+c\agl{p_1}{p_3} \>.
\end{equation}
We remark that this is purely a convention that we decided to adopt for faster typing and has nothing to do with
algebraic properties of the spinors. Thus, upon declaring momentum labels, one can write
\begin{figure}[H]
	\begin{center}
		\makebox[\textwidth]{
			\fbox{\includegraphics[page=1,trim={1.5cm 17.8cm 4cm 9cm},clip]{Example_building_blocks4}}}
	\end{center}
\end{figure}

One should be careful about the delicate interplay between the above defined ``linearity convention" and negative momenta.

\subsection{Chains}

All four of the possible invariants obtained by contracting two spinors with an arbitrary number of momenta, to which we will refer as chains,
are inputted through the single function \texttt{Chain}. This requires as input the type and labels of the spinors which start and end the chain,
and a list of the internal momenta. For example:
\begin{equation}
	\begin{array}{c}
		\langle q_1 \, p_1 \cdots p_{2n} \, q_2 \rangle \hspace{0.3cm} \mapsto \hspace{0.3cm} \texttt{Chain[\$angle,q1,\{p1,...,p2n\},q2,\$angle]} \> , \\
	\end{array}
\end{equation}
and similarly for $[\ldots]$, $\langle \ldots ]$ and $[ \ldots \rangle$.
\begin{figure}[H]
	\begin{center}
		\makebox[\textwidth]{
			\fbox{\includegraphics[page=1,trim={1.5cm 24.5cm 4cm 3.5cm},clip]{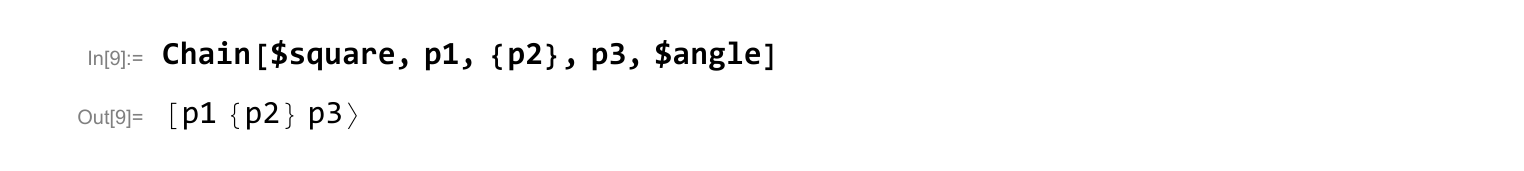}}}
	\end{center}
\end{figure}
Clearly chains are linear with respect to internal momenta, and the external spinors follow the same convention for negative momenta (\fref{eq:negativemom})
as the angle and square brackets do.
\begin{figure}[H]
	\begin{center}
		\makebox[\textwidth]{
			\fbox{\includegraphics[page=1,trim={1.5cm 19.7cm 4cm 6cm},clip]{Example_building_blocks5}}}
	\end{center}
\end{figure}
Once again, for faster typing of expressions we extend the definition \fref{eq:linearitydef} to the external spinors of the chains
as well. This must not be confused with the genuine linearity of the internal momenta.
\begin{figure}[H]
	\begin{center}
		\makebox[\textwidth]{
			\fbox{\includegraphics[page=1,trim={1.5cm 16.6cm 4cm 10.8cm},clip]{Example_building_blocks5}}}
	\end{center}
\end{figure}

\subsection{Uncontracted spinors}\label{sec:uncontracted}

Despite the fact that is generally more convenient to work with Lorentz-invariant quantities such as the spinor brackets and the
Mandelstam invariants, in some instances the use of uncontracted spinors might be required. The direct input of such objects
is accomplished by the functions \texttt{SpinorUndot} and \texttt{SpinorDot}, which require to specify a spinor label, the spinor type ($\lambda$ or reference spinor $\mu$) and a spinor index along with its position.
The spinor type is specified through the use of the protected symbols \texttt{\$lam} and \texttt{\$mu}, and the position specification of the Lorentz index by setting to \texttt{Null} the absent upper or lower index.
\begin{figure}[H]
	\begin{center}
		\makebox[\textwidth]{
			\fbox{\includegraphics[page=1,trim={1.5cm 23cm 4cm 3.5cm},clip]{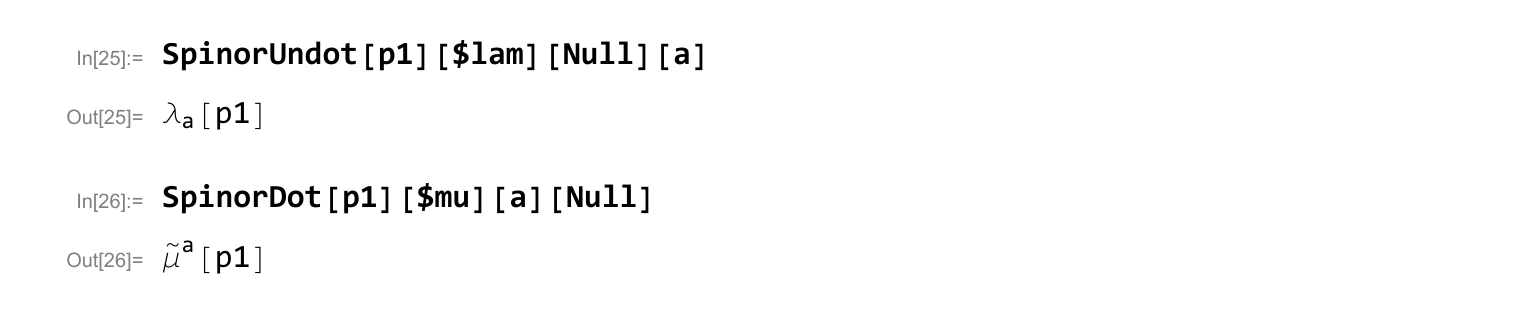}}}
	\end{center}
\end{figure}
We follow the convention that $\lambda_{-p}^\alpha\equiv i\lambda_p^\alpha$ and $\tilde{\lambda}_{-p}^{\dot{\alpha}}\equiv i \tilde{\lambda}_p^{\dot{\alpha}}$ which is automatically
implemented
\begin{figure}[H]
	\begin{center}
		\makebox[\textwidth]{
			\fbox{\includegraphics[page=1,trim={1.5cm 17.5cm 4cm 7.7cm},clip]{Example_building_blocks6}}}
	\end{center}
\end{figure}
Furthermore, just as in the case of the angle and square brackets a pseudo-linearity convention on the momentum labels is in place, allowing for faster
expression input (see \fref{eq:pseudolinearity} and related discussion).
\begin{figure}[H]
	\begin{center}
		\makebox[\textwidth]{
			\fbox{\includegraphics[page=1,trim={1.5cm 5cm 4cm 23cm},clip]{Example_building_blocks6}}}
	\end{center}
\end{figure}
Spinor indices can be raised and lowered through the Levi-Civita tensor:
\begin{figure}[H]
	\begin{center}
		\makebox[\textwidth]{
			\fbox{\includegraphics[page=1,trim={1.5cm 13cm 4cm 13.7cm},clip]{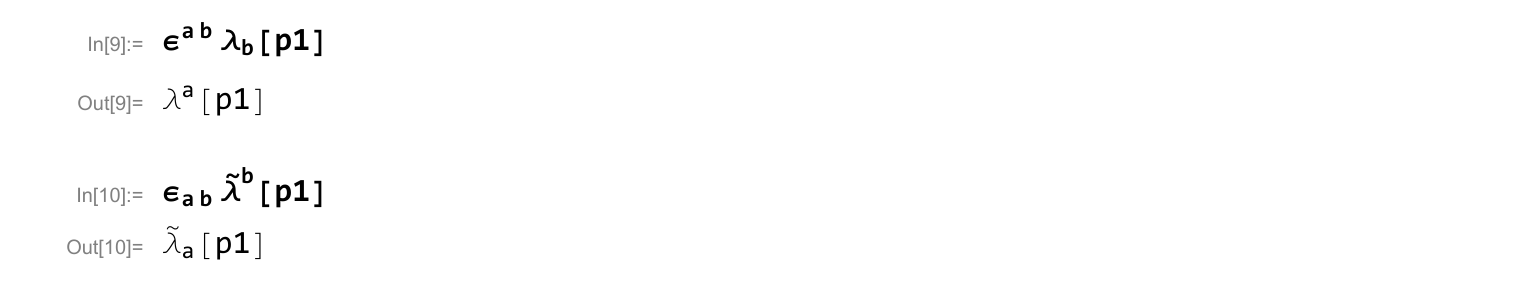}}}
	\end{center}
\end{figure}
In a product of spinors, if possible spinor contractions into invariants are performed:
\begin{figure}[H]
	\begin{center}
		\makebox[\textwidth]{
			\fbox{\includegraphics[page=1,trim={1.5cm 11.3cm 4cm 13.3cm},clip]{Example_building_blocks6}}}
	\end{center}
\end{figure}
Notice also that dotted and undotted indices, belonging to different representations, are not mixed even if the same label is used:
\begin{figure}[H]
	\begin{center}
		\makebox[\textwidth]{
			\fbox{\includegraphics[page=1,trim={1.5cm 8cm 4cm 19.5cm},clip]{Example_building_blocks6}}}
	\end{center}
\end{figure}
Finally, there are a set of objects called \texttt{SpinorUndotBare} and \texttt{SpinorDotBare} which are not physically meaningful
objects but are merely required by the function \texttt{SpinorReplace}, see \fref{sec:spinorreplace}.

\subsection{Reference spinors}

As has already been mentioned in \fref{sec:theory}, it is possible to describe massive particles through the massless spinor helicity formalism, at the price of introducing redundant degrees of freedom in form of reference spinors, see \fref{eq:mommatmassive} which we display here again for convenience:
\begin{equation}
	P_{\alpha\dot{\alpha}}=\lambda_\alpha\tilde{\lambda}_{\dot{\alpha}}+\frac{m^2}{\agl{k}{q}\sqr{q}{k}}\mu_\alpha\tilde{\mu}_{\dot{\alpha}} \>,
\end{equation}
where $P^2=m^2$ and $q$, $k$ are massless with $q_{\alpha\dot{\alpha}}=\lambda_\alpha\tilde{\lambda}_{\dot{\alpha}}$ and $k_{\alpha\dot{\alpha}}=\mu_\alpha\tilde{\mu}_{\dot{\alpha}}$.
We will refer to $k$ as the reference spinor for the massive particle $P$. Since they represent partially redundant degrees of freedom, reference momenta/spinors
get special treatment, both in the uncontracted form of \fref{sec:uncontracted} as well as when they appear inside spinor invariants. Doing so 
allows to simplify expressions by setting the references to convenient values as well as retrace invariants involving massive particles from their spinor components. The way reference spinors are distinguished is by using the $\mu$ label instead of $\lambda$ in the uncontracted form, and by an overbar in the contracted form.
\begin{figure}[H]
	\begin{center}
		\makebox[\textwidth]{
			\fbox{\includegraphics[page=1,trim={1.5cm 23cm 4cm 3.5cm},clip]{Example_building_blocks7}}}
	\end{center}
\end{figure}
Spinor invariants containing reference spinors can be defined using the function \texttt{obar}:
\begin{figure}[H]
	\begin{center}
		\makebox[\textwidth]{
			\fbox{\includegraphics[page=1,trim={1.5cm 20.5cm 4cm 8cm},clip]{Example_building_blocks7}}}
	\end{center}
\end{figure}
\texttt{obar} is linear in declared momenta:
\begin{figure}[H]
	\begin{center}
		\makebox[\textwidth]{
			\fbox{\includegraphics[page=1,trim={1.5cm 17cm 4cm 10.5cm},clip]{Example_building_blocks7}}}
	\end{center}
\end{figure}
Reference spinors are mainly used when dealing with massive states, notice however that also for massless particles reference spinors might come up, for example in the
polarizations, and these can as well be described using the $\mu$ spinors. Since the intrinsic purpose of these built-in objects is the description of massive particles, some care is needed when doing so. In particular, when numerical kinematics is generated the reference spinors of massless particles are automatically initialised to \texttt{Null} and associated masses to zero so that \fref{eq:mommatmassive} reduces to $P^{\dot{\alpha}\alpha}=\tilde{\lambda}^{\dot{\alpha}}\lambda^\alpha$ as it should.

\subsection{Scalar product and Mandelstam invariants}

Scalar products are represented by the protected symbol \texttt{mp}, which has attribute orderless meaning that of course $p_2\cdot p_1=p_1\cdot p_2$,
and is also linear in declared momenta:
\begin{figure}[H]
	\begin{center}
		\makebox[\textwidth]{
			\fbox{\includegraphics[page=1,trim={1.5cm 8.5cm 4cm 17.5cm},clip]{Example_building_blocks7}}}
	\end{center}
\end{figure}
Similarly also the Mandelstam invariants, labelled by \texttt{S}, are orderless but do not present additional properties.
\begin{figure}[H]
	\begin{center}
		\makebox[\textwidth]{
			\fbox{\includegraphics[page=1,trim={1.5cm 6cm 4cm 22.3cm},clip]{Example_building_blocks7}}}
	\end{center}
\end{figure}
The following shorthand notation is available:
\begin{figure}[H]
	\begin{center}
		\makebox[\textwidth]{
			\fbox{\includegraphics[page=1,trim={1.5cm 24cm 4cm 3.8cm},clip]{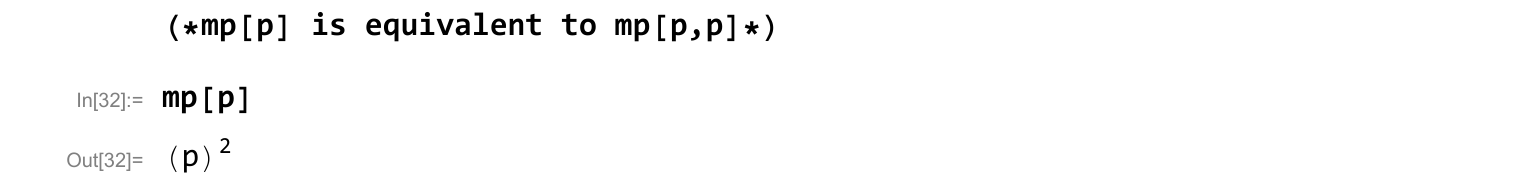}}}
	\end{center}
\end{figure}
When a momentum is declared as massless the associated scalar product is set to zero.
\begin{figure}[H]
	\begin{center}
		\makebox[\textwidth]{
			\fbox{\includegraphics[page=1,trim={1.5cm 20cm 4cm 7.8cm},clip]{Example_building_blocks8}}}
	\end{center}
\end{figure}
In many instances it might be useful to set the expressions of some of the invariants to given values,
this can be achieved with \texttt{SetInvariants}, which takes as input a list of replacements for scalar products and Mandelstam invariants,
and whose output is a list of the invariants whose values have been fixed. For example consider setting $s_{12}\equiv s$ and $p_1\cdot p_3 \equiv t/2$:
\begin{figure}[H]
	\begin{center}
		\makebox[\textwidth]{
			\fbox{\includegraphics[page=1,trim={1.5cm 14cm 4cm 11cm},clip]{Example_building_blocks8}}}
	\end{center}
\end{figure}
Clearly, if $p_i$ and $p_j$ are massless $s_{ij}=2 p_i\cdot p_j$, which is account for by \texttt{SetInvariants}:
\begin{figure}[H]
	\begin{center}
		\makebox[\textwidth]{
			\fbox{\includegraphics[page=1,trim={1.5cm 7cm 4cm 17cm},clip]{Example_building_blocks8}}}
	\end{center}
\end{figure}
If \texttt{SetInvariants} is called without arguments it simply displays the list of already fixed values, also invariants are cleared with \texttt{ClearInvariants}, which takes as input the invariants to be cleared. If no argument is given to \texttt{ClearInvariants} all the invariants are cleared.
\begin{figure}[H]
	\begin{center}
		\makebox[\textwidth]{
			\fbox{\includegraphics[page=1,trim={1.5cm 2.8cm 4cm 24cm},clip]{Example_building_blocks8}}}
	\end{center}
\end{figure}
Notice that \texttt{SetInvariants} does not require the invariants to be set to constants:
\begin{figure}[H]
	\begin{center}
		\makebox[\textwidth]{
			\fbox{\includegraphics[page=2,trim={1.5cm 24cm 4cm 3.5cm},clip]{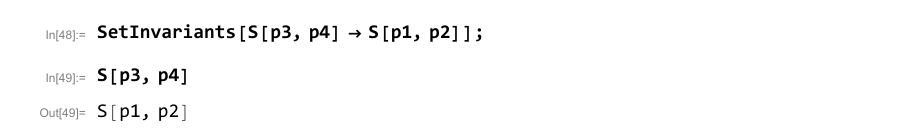}}}
	\end{center}
\end{figure}

\subsection{Vector quantities}

The package also features a limited number of vector quantities which often appear in amplitudes calculations. These include uncontracted momentum vectors, polarization vectors, the flat space metric tensor $\eta$, the Kronecker delta and the vector of Pauli matrices\footnote{These are only required as a placeholders inside Chain expressions.}. The polarizations also present an input form stripped of the Lorentz index called \texttt{PolarBare}, which is required when the polarizations appear inside scalar products for example.
\begin{figure}[H]
	\begin{center}
		\makebox[\textwidth]{
			\fbox{\includegraphics[page=1,trim={1.5cm 15.5cm 4cm 5cm},clip]{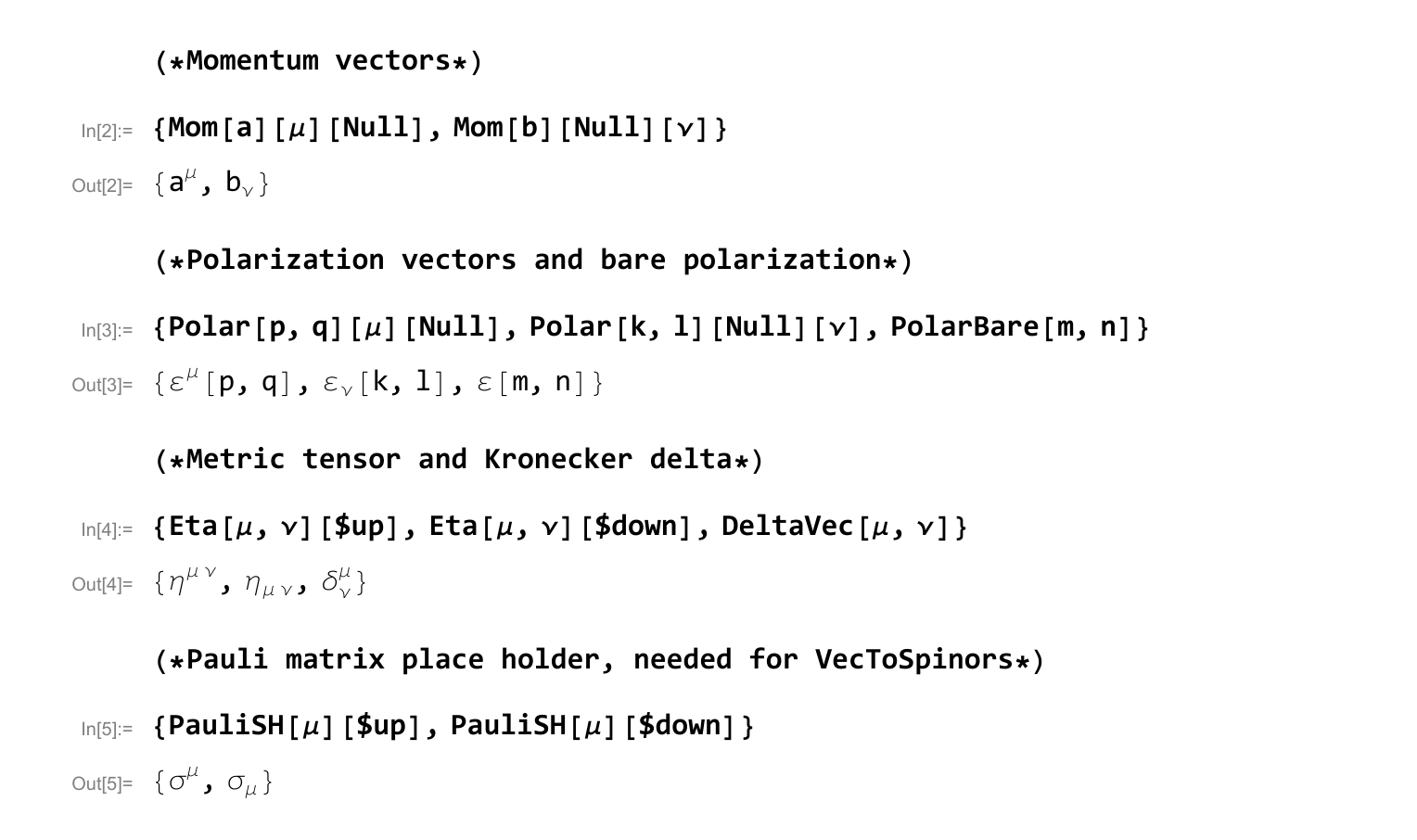}}}
	\end{center}
\end{figure}
All the vector quantities enjoy automatic contraction properties among each other.
\begin{figure}[H]
	\begin{center}
		\makebox[\textwidth]{
			\fbox{\includegraphics[page=1,trim={1.5cm 7cm 4cm 15.5cm},clip]{Example_Vector_quantities}}}
	\end{center}
\end{figure}
These contraction properties include the trace of the Kronecker delta, whose value can be modified through the option \texttt{\$DeltaVecDim} of \texttt{DeltaVec}.
\begin{figure}[H]
	\begin{center}
		\makebox[\textwidth]{
			\fbox{\includegraphics[page=2,trim={1.5cm 16cm 4cm 7.5cm},clip]{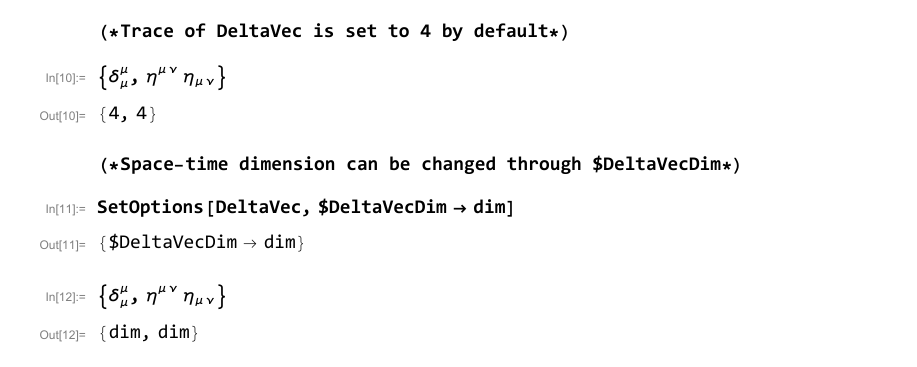}}}
	\end{center}
\end{figure}

\subsection{Clearing everything, NewProcess}

It is possible to reset all the given definitions relating to the building blocks presented in this whole section, without the need of quitting the \texttt{Mathematica}
kernel. This is achieved through \texttt{NewProcess} which clears the values of the invariants, the list of momenta declared as massless and the list of labels declared to be momenta.
\begin{figure}[H]
	\begin{center}
		\makebox[\textwidth]{
			\fbox{\includegraphics[page=2,trim={1.5cm 16cm 4cm 7cm},clip]{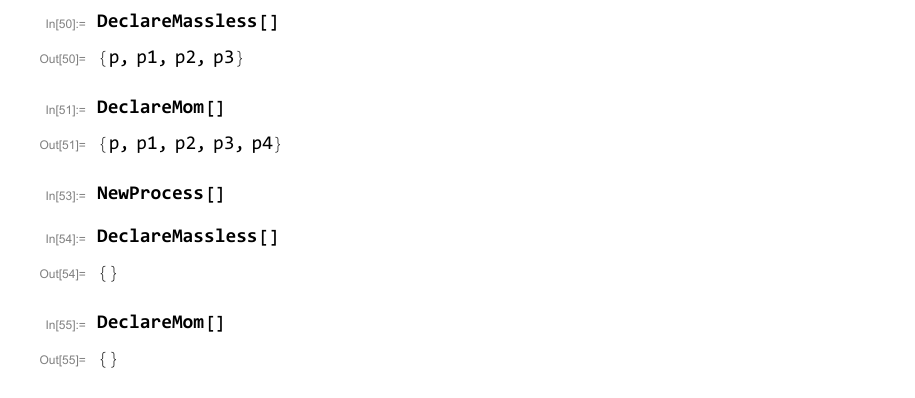}}}
	\end{center}
\end{figure}

\section{Manipulation of analytic expressions}
In this section we describe the manipulation of given analytic expressions defined in terms of the building blocks introduced so far.

\subsection{ToChain, contraction of spinor products into chains}

When dealing with the product of multiple angle and square brackets it is possible to contract them into continuos chains by means of the replacement
\begin{equation}
	\cdots|p\rangle[p|\cdots \mapsto \cdots |p|\cdots \>,
\end{equation}
this operation is achieved through \texttt{ToChain}. The contraction of spinor products into chains is not unique, consider for example the following
\begin{equation}
	\agl{1}{2}\sqr{2}{5}\sqr{2}{3}\agl{3}{4} = \begin{cases}
		\langle 1 \, 2\, 5][ 2 \, 3 \, 4 \rangle \\
		\langle 1 \, 2 \, 3 \, 4 \rangle \sqr{2}{5}
	\end{cases}
\end{equation}
both are perfectly valid contractions but depending on the situation one or the other may be preferred. \texttt{ToChain} allows for the option \texttt{ChainSelection} which provides some degree of control over how chain contractions are performed:
\begin{itemize}
	\item ``MostTraces'' selects the contraction which maximises the number of Dirac traces in the expression, in other words the number of chains of the type $\langle p \cdots p ]$
	\item ``LongestChain'' tries to find the longest continuos contraction into a chain
	\item ``ShortestChain'' tries to find the shortest contractions where still as many spinor brackets as possible are contracted into a chain\footnote{Of course without the latter requirement the shortest contraction would simply be no contraction at all.}
	\item ``RandomChain'' is the default argument and provides a contraction based on no specific criterion.
\end{itemize}
In order for any of the first three criteria to be applied it is first necessary to find all possible contractions and then select the appropriate one. The price to pay comes in terms of computational efficiency: since of course the more spinor invariants there are the more contractions are possible, the more computationally expensive it becomes to find all the chain contractions. There are however situations in which one might only be interested in getting one possible contraction but not be concerned about which one. This is the case for example when one considers a helicity-neutral structure, such as those often appearing in generalised unitarity calculations upon factoring out an appropriate overall helicity factor. In these situations we recommend the use of the default option ``RandomChain'' which is based on an algorithm designed to simply return one possible contraction without classifying them all, and is thus much faster than the other criteria.

\begin{figure}[H]
	\begin{center}
		\makebox[\textwidth]{
			\fbox{\includegraphics[page=1,trim={1.5cm 12cm 4cm 4.5cm},clip]{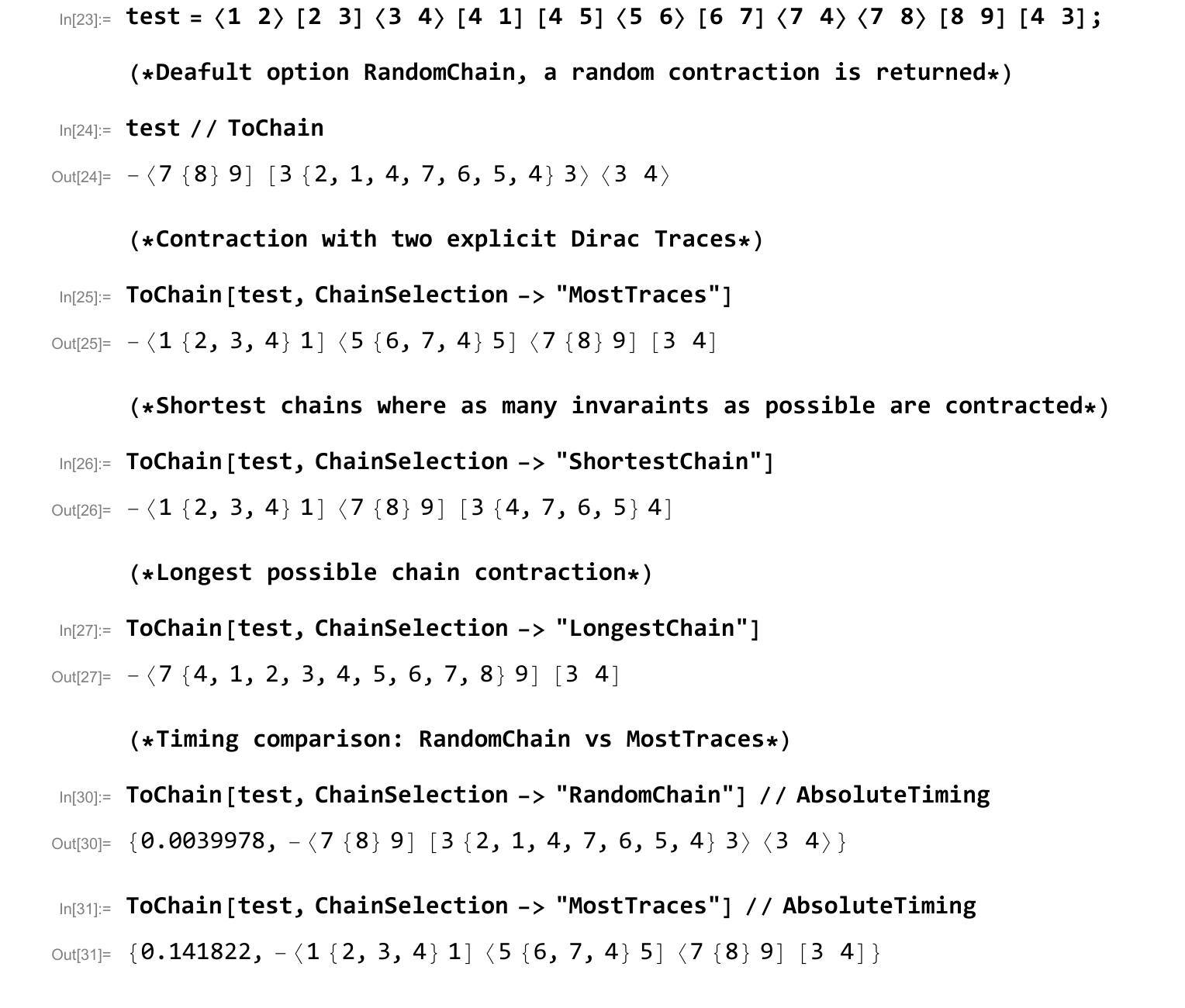}}}
	\end{center}
\end{figure}

\subsection{ChainToSpinor}

The function \texttt{ChainToSpinor} performs the inverse operation of \texttt{ToChain} by splitting chains into spinor brackets. Notice that, while the contraction into chains is always possible, splitting a chain into spinor brackets is only allowed if the involved particles are massless, which needs to be declared through \texttt{DeclareMassless}.
\begin{figure}[H]
	\begin{center}
		\makebox[\textwidth]{
			\fbox{\includegraphics[page=1,trim={1.5cm 19cm 4cm 4.5cm},clip]{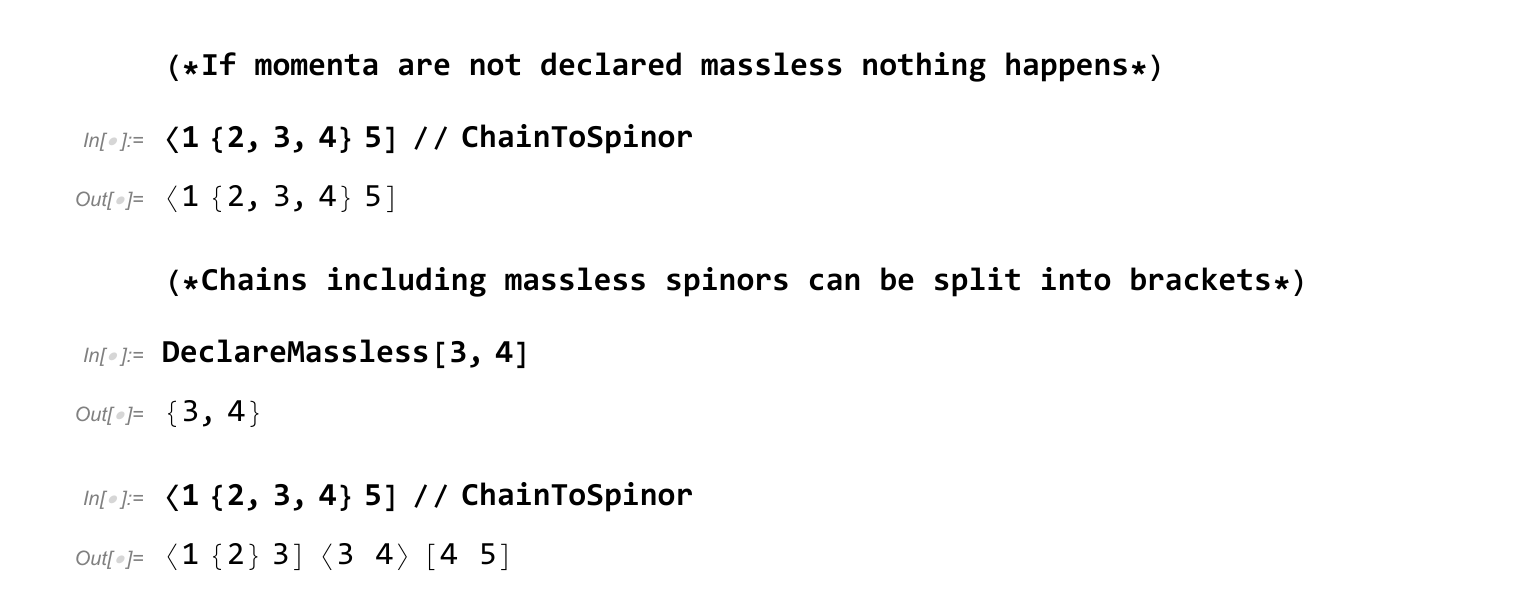}}}
	\end{center}
\end{figure}

\subsection{ChainMomentumCon}

The function \texttt{ChainMomentumCon} allows to apply momentum conservation identities at the chain level without affecting uncontracted spinor brackets. If momentum labels are declared, linearity is as always applied.
\begin{figure}[H]
	\begin{center}
		\makebox[\textwidth]{
			\fbox{\includegraphics[page=1,trim={1.5cm 18.5cm 4cm 4.5cm},clip]{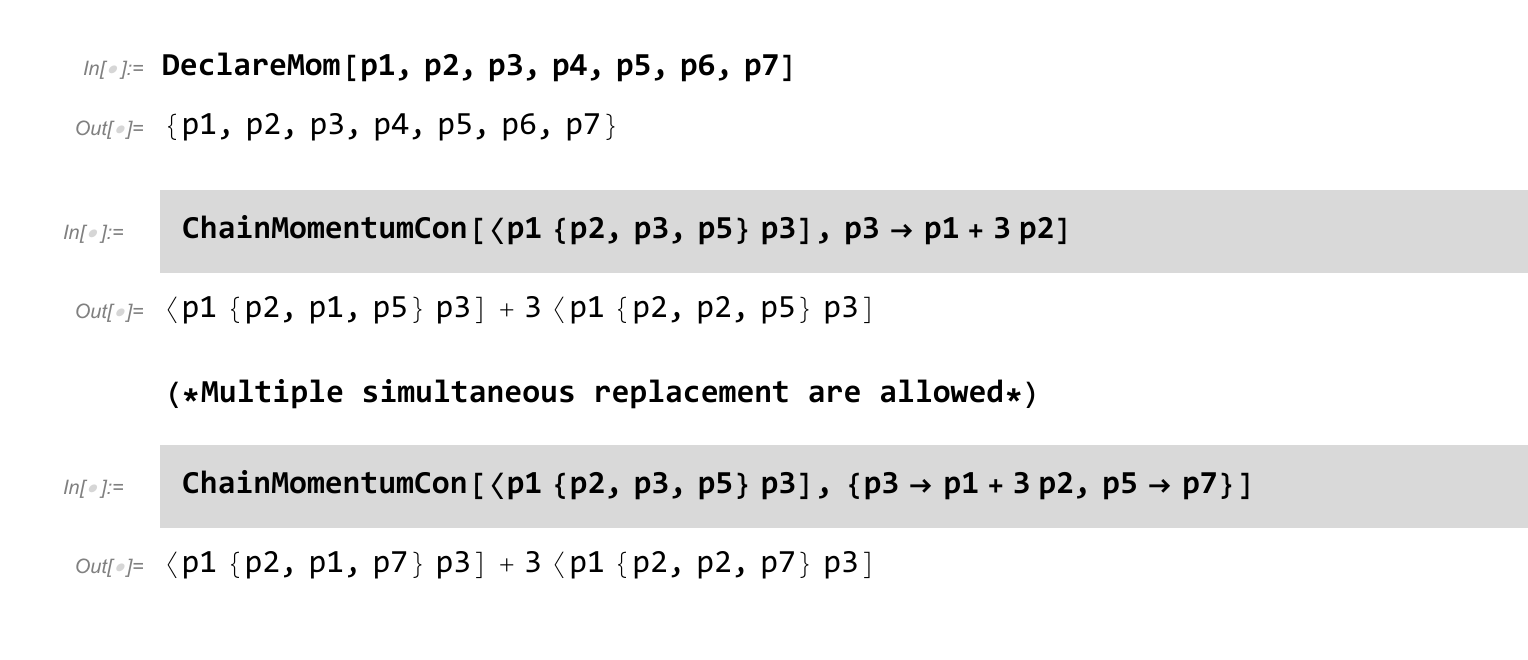}}}
	\end{center}
\end{figure}
In the particular case in which the momentum which we want to replace appears as the extrema of a Dirac trace chain, the chain is automatically rearranged so to allow for the application of momentum conservation. The rearrangement of the chain into a different Dirac trace is only possible if at least one of the momenta inside the chain is massless.
\begin{figure}[H]
	\begin{center}
		\makebox[\textwidth]{
			\fbox{\includegraphics[page=1,trim={1.5cm 11cm 4cm 12cm},clip]{Example_ChainMomCon}}}
	\end{center}
\end{figure}

\subsection{ChainSort}

The function \texttt{ChainSort} allows to sort the momenta present in contracted chains through the use of the Clifford algebra identity of the $\sigma$ matrices $\{\sigma^\mu, \bar{\sigma}^\nu \}=2\eta^{\mu \nu}$. This procedure can prove quite useful in simplification of expressions because it allows to reduce all chains to a common canonically ordered form. Notice that the sorting procedure is customizable, if an optional argument consisting of a list of labelles is passed to the function, this is used to sort the chains instead of the default ordering provided by \texttt{Mathematica}.
\begin{figure}[H]
	\begin{center}
		\makebox[\textwidth]{
			\fbox{\includegraphics[page=1,trim={1.5cm 15.5cm 4cm 4.5cm},clip]{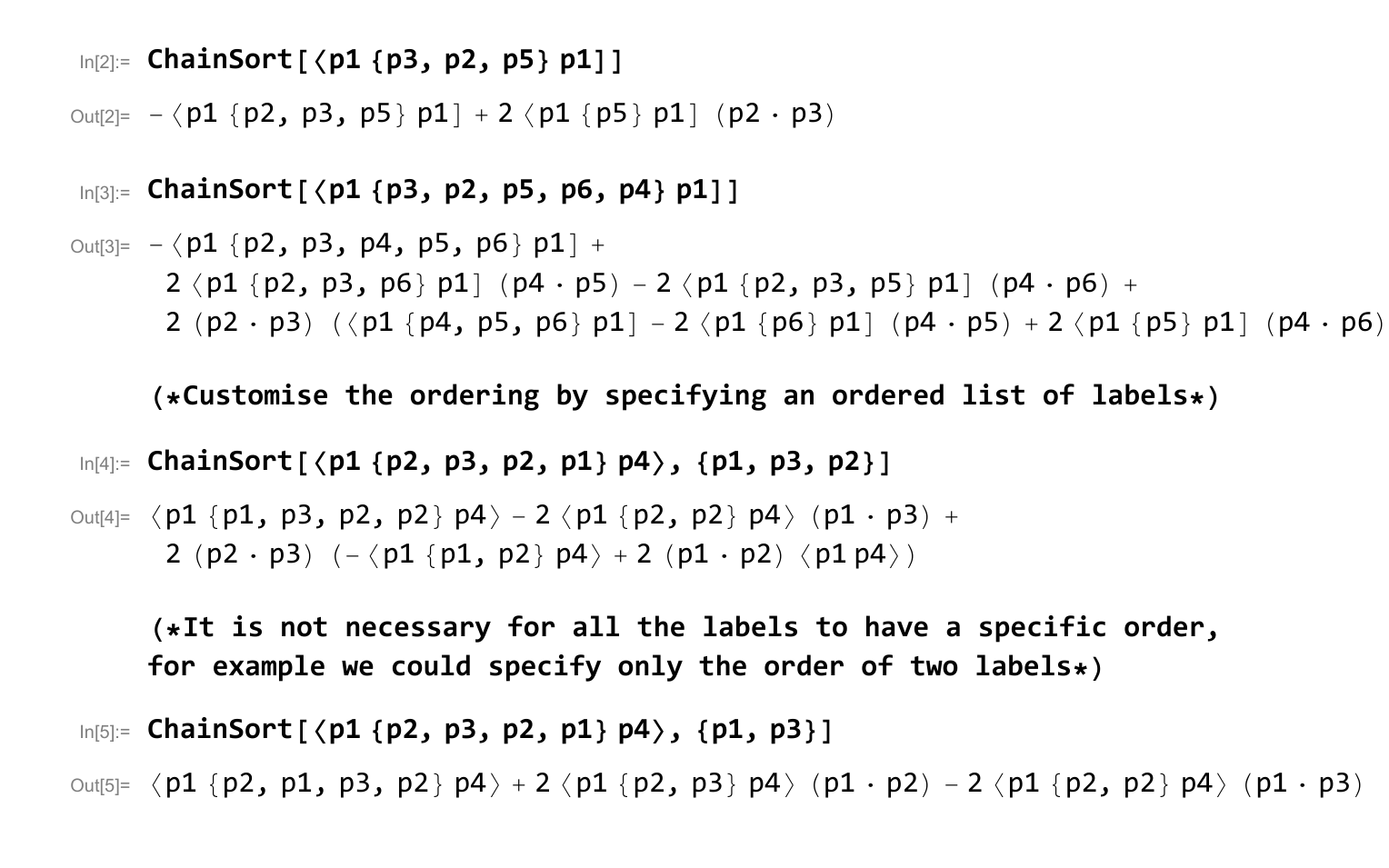}}}
	\end{center}
\end{figure}
The extrema of the chain are also included in the sorting, and in the case of a Dirac trace (where the extrema can often be arbitrarily chosen) these can also mix with internal massless momentum labels.
\begin{figure}[H]
	\begin{center}
		\makebox[\textwidth]{
			\fbox{\includegraphics[page=1,trim={1.5cm 4.5cm 4cm 14.5cm},clip]{Example_ChainSort}}}
	\end{center}
\end{figure}

\subsection{ChainSimplify}

While sorting the momenta inside chains may lead to simplifications, a more powerfull simplification algorithm is available to the user in form of \texttt{ChainSimplify}. This function, which takes as single input the expression to be simplified, makes use once again of $\{\sigma^\mu, \bar{\sigma}^\nu \}=2\eta^{\mu \nu}$ to reshuffle momenta in such a way that chains in the input reduce to shorter chains and scalar products. Take as an example
\begin{equation}
	\langle q \, p \, q \, k \rangle = 2 p\cdot q \, \agl{q}{k} - \langle q \, q \, p \, k \rangle = 2 p\cdot q \,  \agl{q}{k} \> .
\end{equation}
We stress that even simplifications which do not require reshuffling of the chains are performed by \texttt{ChainSimplify}, this includes for example the trivial $\langle q \, q \, p \, k \rangle = 0$.
\begin{figure}[H]
	\begin{center}
		\makebox[\textwidth]{
			\fbox{\includegraphics[page=1,trim={1.5cm 20.5cm 4cm 4.75cm},clip]{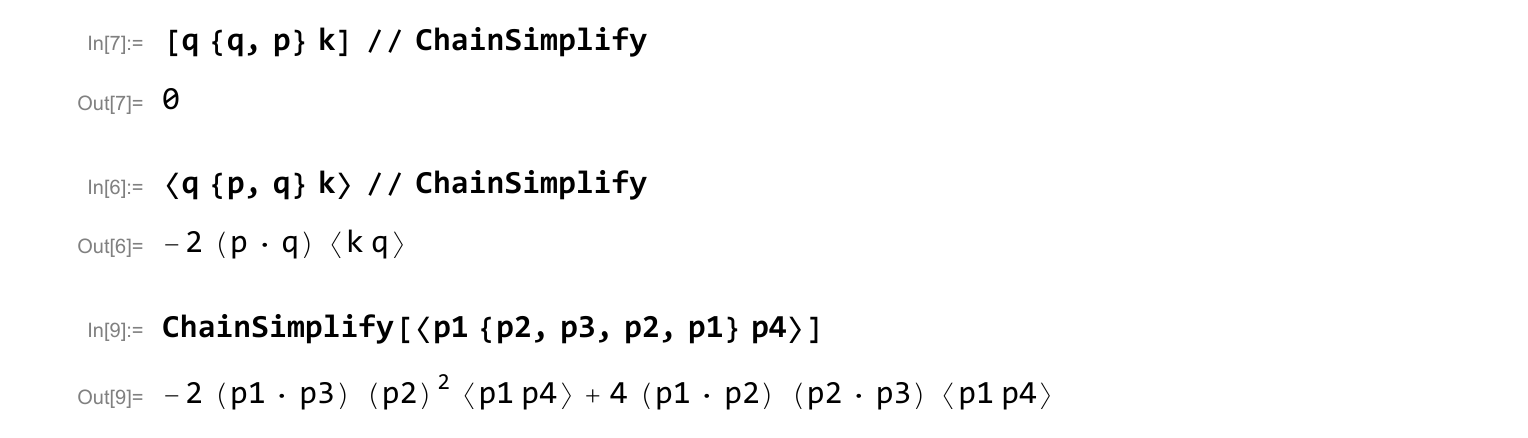}}}
	\end{center}
\end{figure}

Unless it is explicitly requested through the option \texttt{ReduceBySorting} momenta in the chains are not sorted, since a priori sorting does not guarantee a simpler final expression. By simply setting \texttt{ReduceBySorting -> True} standard ordering of the momenta is applied, while specifying a list allows to specify a custom ordering.

\begin{figure}[H]
	\begin{center}
		\makebox[\textwidth]{
			\fbox{\includegraphics[page=1,trim={1.5cm 12cm 4cm 10cm},clip]{Example_ChainSimplify}}}
	\end{center}
\end{figure}

The option \texttt{ReduceBySorting} under the hood calls the function \texttt{ChainSort}, thus the same rules apply, for example in the case of Dirac traces:
\begin{figure}[H]
	\begin{center}
		\makebox[\textwidth]{
			\fbox{\includegraphics[page=1,trim={1.5cm 6cm 4cm 18.5cm},clip]{Example_ChainSimplify}}}
	\end{center}
\end{figure}
It is also possible to account for momentum conservation directly inside \texttt{ChainSimplify} through the option \texttt{MomCon}. This is equivalent to first applying the function \texttt{ChainMomentumCon} and then \texttt{ChainSimplify} right after. Notice that if a momentum conservation relation is given, it will be applied, independently of whether the final expression is simpler or not.
\begin{figure}[H]
	\begin{center}
		\makebox[\textwidth]{
			\fbox{\includegraphics[page=2,trim={1.5cm 24.5cm 4cm 2.5cm},clip]{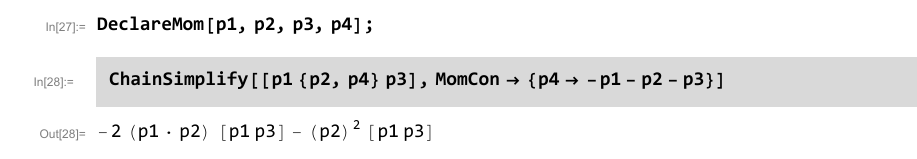}}}
	\end{center}
\end{figure}

\subsection{ToTrace}
As a final application related to chain manipulation, we have the function \texttt{ToTrace}. This function allows to convert chains of the form $\langle q \cdots q ]$ and $[q \cdots q \rangle$, which correspond to Dirac traces with a $\gamma_5$ insertion, into products of scalar products and Levi-Civita tensors dotted into momenta, through the identity
\begin{equation}
\def\arraystretch{2}
\begin{array}{l}
\langle q \; p_1 \ldots  p_{2n+1} \; q ] = \text{Tr}\left[ \left(\frac{\mathbf{1}-\gamma_5}{2}\right)\gamma_{\mu}\ldots \gamma_{\nu}\right] \; p_1^{\mu}\ldots p_{2n+1}^{\nu}\> , \\
\> [ q \; p_1 \ldots  p_{2n+1} \; q\rangle = \text{Tr}\left[ \left(\frac{\mathbf{1}+\gamma_5}{2}\right)\gamma_{\mu}\ldots \gamma_{\nu}\right] \; p_1^{\mu}\ldots p_{2n+1}^{\nu} \> .
\end{array}
\end{equation}
In the package Levi-Civita tensors are repesented by the object \texttt{epsSH}, where \texttt{epsSH[p1,p2,p3,p4]} =$\epsilon_{\mu_1,\mu_2,\mu_3,\mu_4}p_1^{\mu_1}p_2^{\mu_2}p_3^{\mu_3}p_4^{\mu_4}$.
\begin{figure}[H]
	\begin{center}
		\makebox[\textwidth]{
			\fbox{\includegraphics[page=1,trim={2.5cm 23cm 3cm 4.5cm},clip]{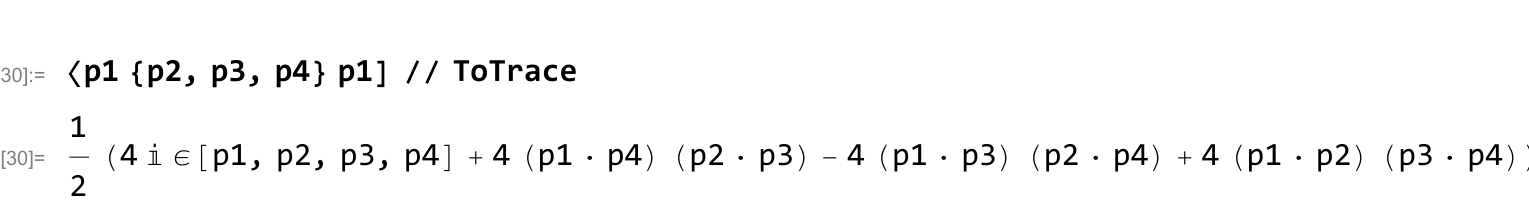}}}
	\end{center}
\end{figure}
The function allows for the option \texttt{EpsilonSimplify} which by default is set to \texttt{None}. It can take the following string values:
\begin{itemize}
\item \texttt{ReduceEven}, this option takes into account the fact that the product of two $\epsilon$ tensors can be rewritten as a product of Kronecker deltas, and thus in our case where the $\epsilon$ tensors are always completely contracted into a set of momenta one gets
\begin{equation}\label{eq:evenepsilon}
\epsilon(p_1,p_1,p_3,p_4)\epsilon(q_1,q_2,q_3,q_4)=
\begin{vmatrix}
(p_1 \cdot q_1) & \cdots & (p_1\cdot q_4) \\
\vdots & & \vdots \\
(p_4 \cdot q_1) & \cdots & (p_4 \cdot q_4)
\end{vmatrix} \> ,
\end{equation}
which is the Gram determinant, and where $\epsilon(p_1,p_1,p_3,p_4)$ is the Levi-Civita tensor contracted into the four momenta $p_i$. Setting \texttt{EpsilonSimplify->ReduceEven} will thus convert all products of an even number of epsilon tensors in the output into products of scalar products.
\item \texttt{KillOdd}, this option, beyond applying the relation \eqref{eq:evenepsilon}, also sets to zero all the combinations of an odd number of epsilon tensors. This might be useful for example when dealing with a four-particle process, where thanks to momentum conservation $\epsilon[p_1,p_2,p_3,p_4] = -\epsilon[p_1,p_2,p_3,p_1] -\epsilon[p_1,p_2,p_3,p_2]-\epsilon[p_1,p_2,p_3,p_3] = 0$. 
\end{itemize}
\begin{figure}[H]
	\begin{center}
		\makebox[\textwidth]{
			\fbox{\includegraphics[page=1,trim={3cm 8cm 2cm 7.5cm},clip]{Example_ToTrace}}}
	\end{center}
\end{figure}

The function \texttt{ToTrace} uses the auxiliary functions \texttt{TrG} and \texttt{TrG5} which separately perform the computations of $\text{Tr}\left[\gamma_{\mu}\ldots \gamma_{\nu}\right] \; p_1^{\mu}\ldots p_{2n+1}^{\nu}$ and $\text{Tr}\left[ \gamma_5\gamma_{\mu}\ldots \gamma_{\nu}\right] \; p_1^{\mu}\ldots p_{2n+1}^{\nu}$ respectively.
\begin{figure}[H]
	\begin{center}
		\makebox[\textwidth]{
			\fbox{\includegraphics[page=1,trim={1.5cm 4cm 4cm 22.5cm},clip]{Example_ToTrace}}}
	\end{center}
\end{figure}

\subsection{SchoutenSimplify}

A key component in the simplification of spinor expressions is the application of Schouten identities, which we recall here for the reader's convenience:
\begin{equation}
	\agl{i}{j}\langle k | +\agl{j}{k}\langle i | + \agl{k}{i}\langle j |=0 \>, \hspace{1.5cm}
	\sqr{i}{j}[k | +\sqr{j}{k}[i | + \sqr{k}{i}[j |=0 \>.
\end{equation}

Application of these identities is performed through the function \texttt{SchoutenSimplify}, which takes as single input the expression to be simplified.

\begin{figure}[H]
	\begin{center}
		\makebox[\textwidth]{
			\fbox{\includegraphics[page=1,trim={1.5cm 18.5cm 4cm 4.5cm},clip]{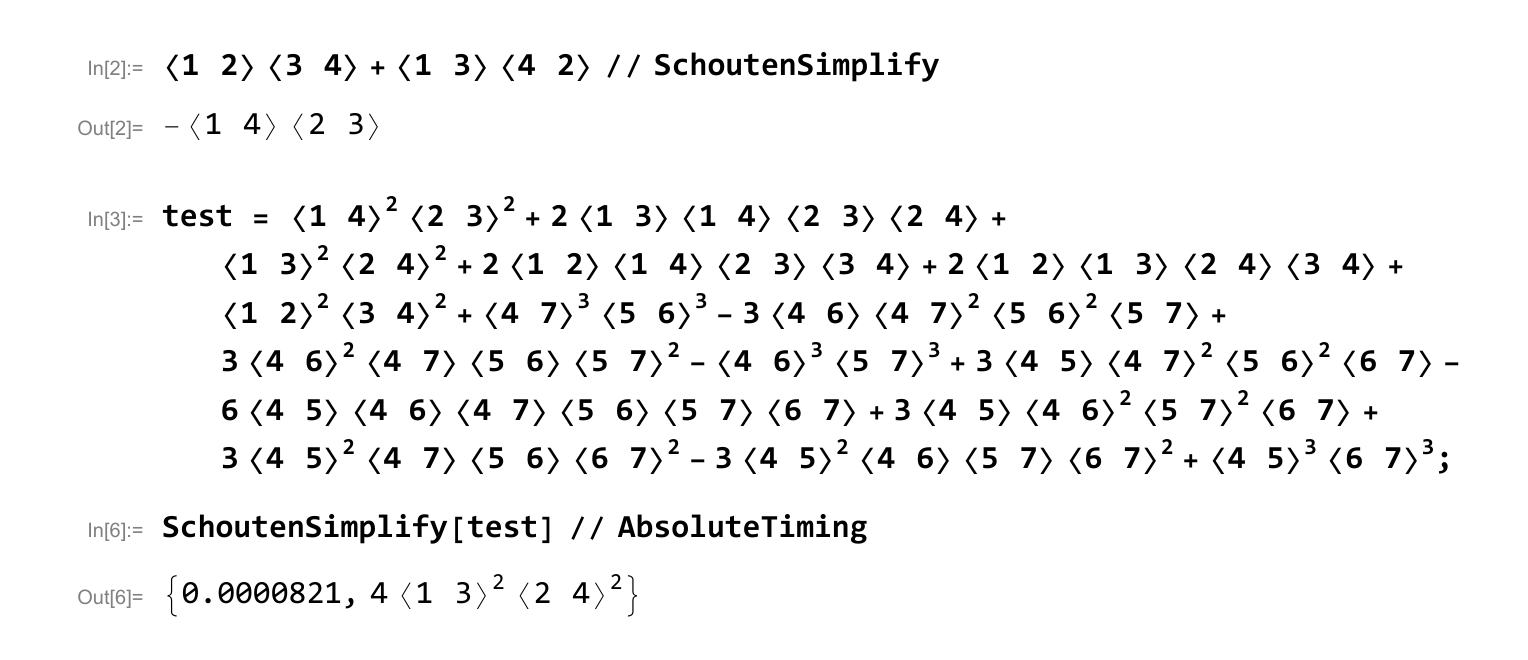}}}
	\end{center}
\end{figure}
\texttt{SchoutenSimplify} under the hood runs a modified version of \texttt{Mathematica}'s function \texttt{FullSimplify} and thus allows for the same options, being also subject to the same limitations. 

\begin{figure}[H]
	\begin{center}
		\makebox[\textwidth]{
			\fbox{\includegraphics[page=1,trim={1.5cm 12.5cm 4cm 12.5cm},clip]{Example_SchoutenSimplify}}}
	\end{center}
\end{figure}

\subsection{SpinorReplace}\label{sec:spinorreplace}

Replacing spinor labels inside an expression might seem a simple enough task, however it can clearly not be accomplished by a mere $exp \, /. \, p_{old} \to p_{new}$. Spinor replacements are performed through the function \texttt{SpinorReplace[expression,rules]}, which acts on angle and square brackets as well as chain extrema. We stress that
\begin{itemize}
	\item replacement rules are given in terms of bare spinors, which do not have explicit indices since the replacement will be applied independently of the index structure. Bare spinors have the same ``linearity'' properties with respect to declared momenta as all the other building blocks.
	\begin{figure}[H]
		\begin{center}
			\makebox[\textwidth]{
				\fbox{\includegraphics[page=1,trim={1.5cm 21cm 4cm 4.5cm},clip]{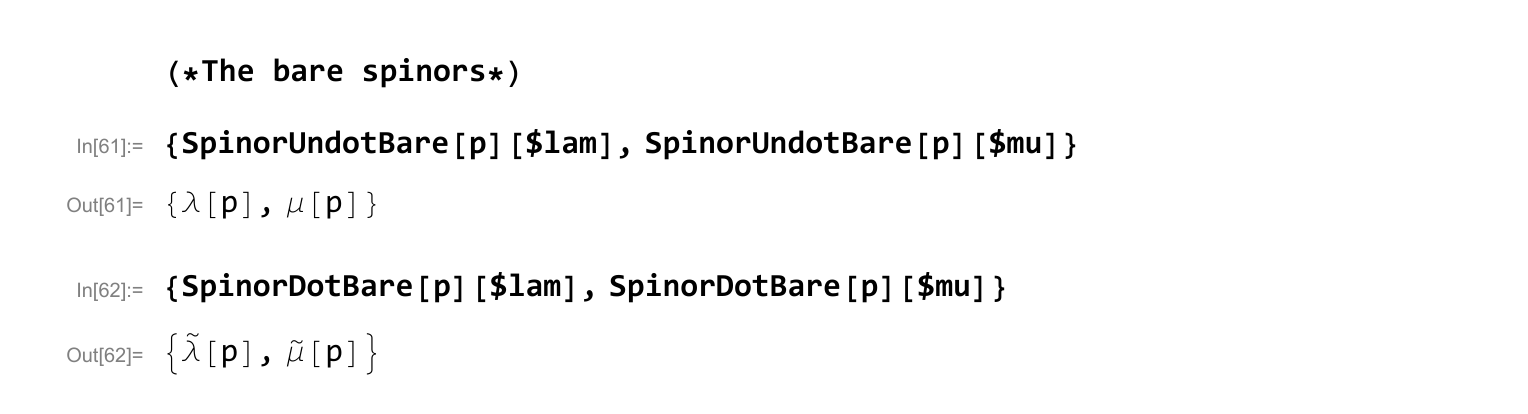}}}
		\end{center}
	\end{figure}
	\item replacements can involve any type of linear combinations of spinors. Notice that 
	
	\texttt{SpinorDotPure[p1]->SpinorDotPure[p2+z*p3]} is equivalent to 
	
	\texttt{SpinorDotPure[p1]->SpinorDotPure[p2]+z*SpinorDotPure[p3]} if \texttt{p1,p2,p3} have previously been declared as momentum labels through \texttt{DeclareMom}.
	\begin{figure}[H]
		\begin{center}
			\makebox[\textwidth]{
				\fbox{\includegraphics[page=1,trim={1.5cm 10cm 4cm 9.5cm},clip]{Example_SpinorReplace}}}
		\end{center}
	\end{figure}
	\item replacements can involve linear combinations of spinors dotted into momentum matrices simply through a standard dot, without the need to declare the momenta previously
	\begin{figure}[H]
		\begin{center}
			\makebox[\textwidth]{
				\fbox{\includegraphics[page=1,trim={1.5cm 4cm 4cm 20.5cm},clip]{Example_SpinorReplace}}}
		\end{center}
	\end{figure}
	\item Replacements involving reference spinors work the same way, just by using bare $\mu$ spinors insteam of bare $\lambda$ spinors.
	\begin{figure}[H]
		\begin{center}
			\makebox[\textwidth]{
				\fbox{\includegraphics[page=2,trim={1.5cm 23cm 4cm 4cm},clip]{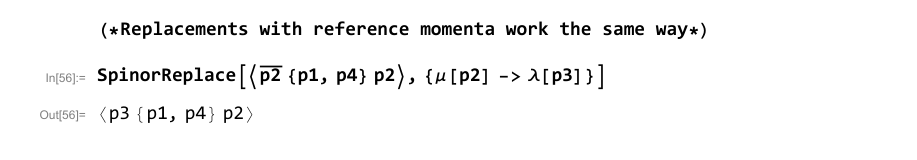}}}
		\end{center}
	\end{figure}
	\end{itemize}

\subsection{SpinorDerivative}

Another interesting feature available in the package is the function \texttt{SpinorDerivative}, which allows to take the derivative of spinorial expressions with respect to the spinors themselves. This function takes two inputs, the expression to be differentiated and the spinor with respect to which to differentiate.
\begin{figure}[H]
	\begin{center}
		\makebox[\textwidth]{
			\fbox{\includegraphics[page=1,trim={2.5cm 21cm 2.5cm 4.5cm},clip]{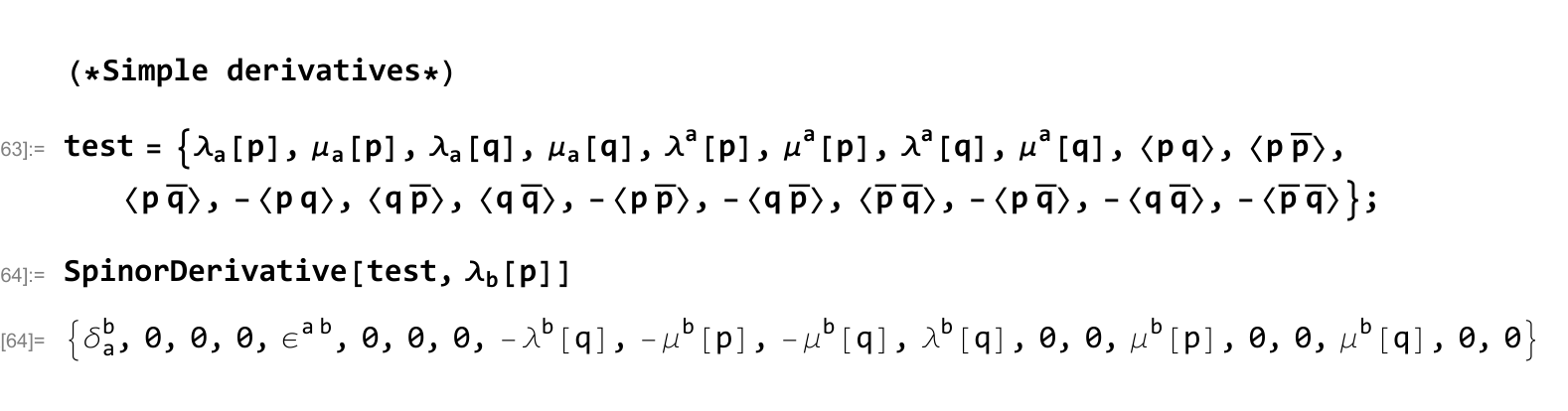}}}
	\end{center}
\end{figure}
Giving a lits or a products of spinor as second arguments wil perform sequential derivatives with respect to all the specified spinors:
\begin{figure}[H]
	\begin{center}
		\makebox[\textwidth]{
			\fbox{\includegraphics[page=1,trim={1.5cm 15.5cm 4cm 9.5cm},clip]{Example_SpinorDerivative}}}
	\end{center}
\end{figure}
\texttt{SpinorDerivative} can differentiate any arbitrary function of the spinors.
\begin{figure}[H]
	\begin{center}
		\makebox[\textwidth]{
			\fbox{\includegraphics[page=1,trim={1.5cm 6cm 4cm 15cm},clip]{Example_SpinorDerivative}}}
	\end{center}
\end{figure}

\subsection{ToMandelstam}

The function \texttt{ToMandelstam} converts products of spinor brackets as well as scalar products into Mandelstam invariants, in other words it performs the replacement $\agl{i}{j}\sqr{j}{i}, 2\, p_i \cdot p_j \mapsto s_{ij}$, as long as the momenta $p_i$ and $p_j$ have been declared as massless.
\begin{figure}[H]
	\begin{center}
		\makebox[\textwidth]{
			\fbox{\includegraphics[page=1,trim={1.5cm 15.5cm 4cm 4.5cm},clip]{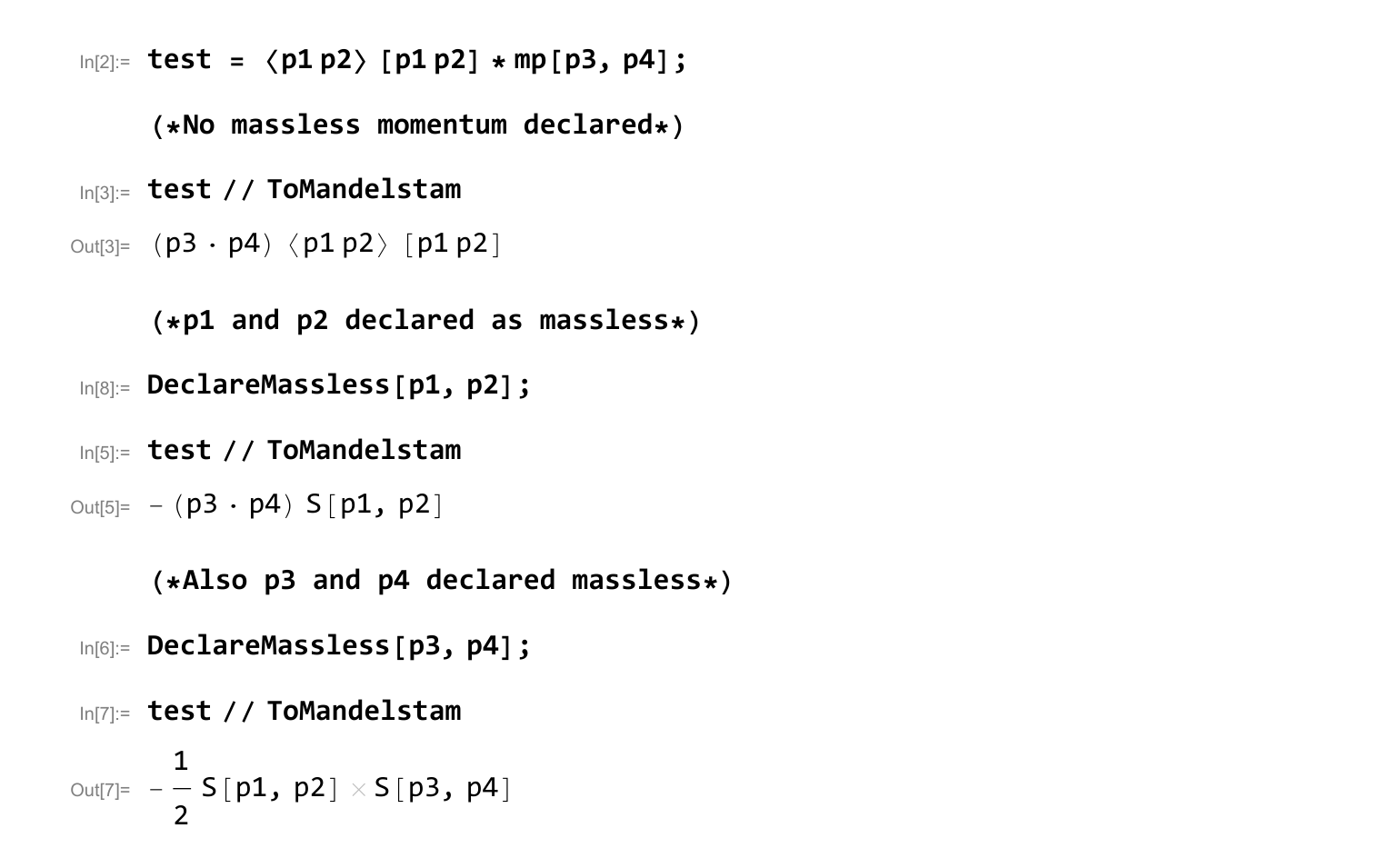}}}
	\end{center}
\end{figure}

\subsection{VecToSpinors}
The function \texttt{VecToSpinors} allows to convert uncontracted vector quantities in the input to to spinor expressions, in particular to spinor chains with an uncontracted Pauli matrix. The function requires as input argument the expression to be converted, and optionally two lists of momentum labels for the positive and negative helicity states. The latter are only required when polarization vectors are included in the input. Notice that currently this mapping is available only for massless momenta.
\begin{figure}[H]
	\begin{center}
		\makebox[\textwidth]{
			\fbox{\includegraphics[page=1,trim={1.5cm 14.5cm 4cm 5cm},clip]{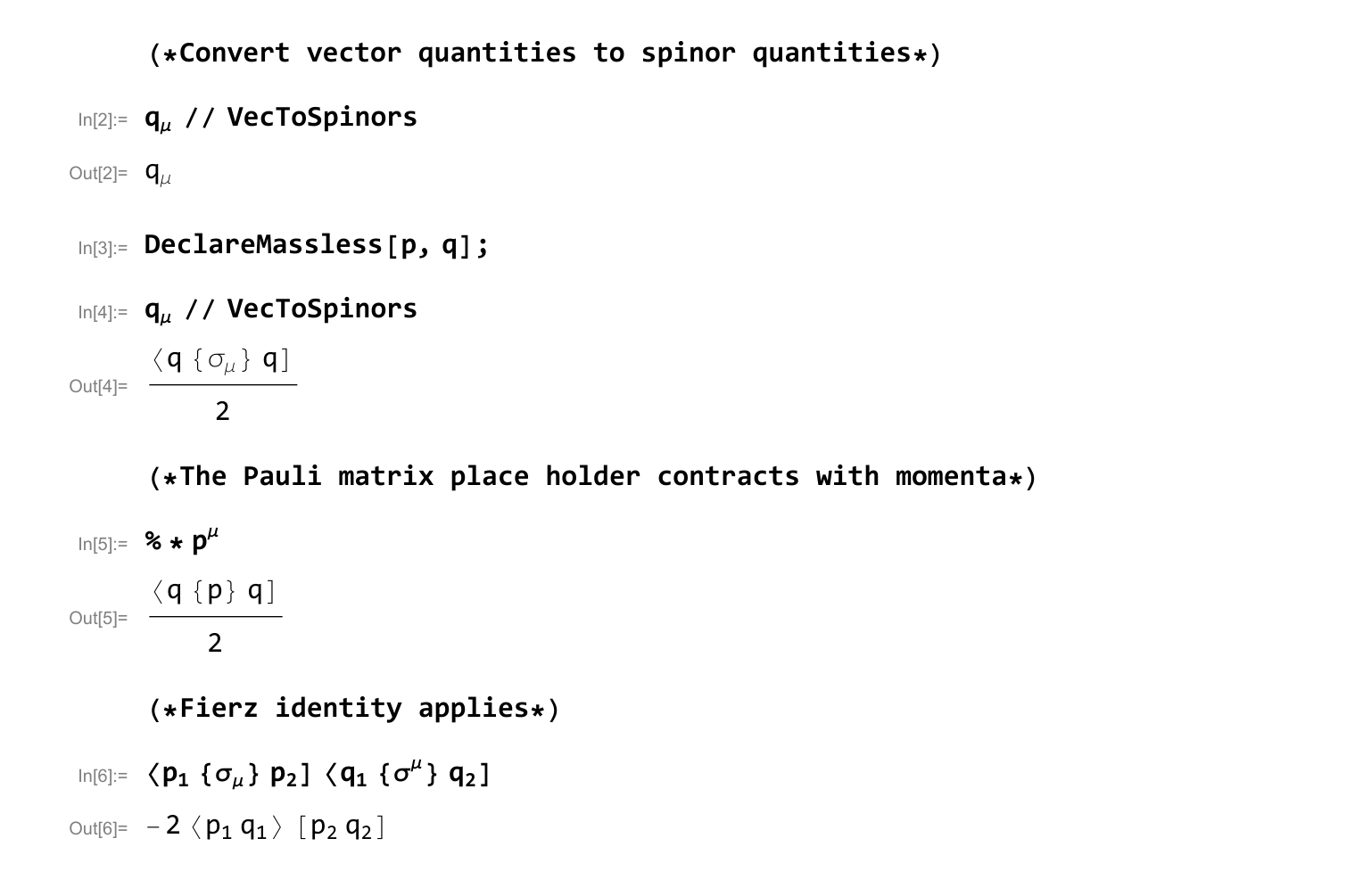}}}
	\end{center}
\end{figure}
Below, specification of the helicity for polarization vectors.
\begin{figure}[H]
	\begin{center}
		\makebox[\textwidth]{
			\fbox{\includegraphics[page=1,trim={1.5cm 5cm 4cm 16.5cm},clip]{Example_VecToSpinors}}}
	\end{center}
\end{figure}

\subsection{MpToSpinors}

The function \texttt{MpToSpinors} converts scalar products into spinor products, it takes as argument the input expression to be converted as well as two optional arguments being lists of momentum labels identifying positive and negative helicity states. Once again, notice that the current implementation only works for massless states.
\begin{figure}[H]
	\begin{center}
		\makebox[\textwidth]{
			\fbox{\includegraphics[page=1,trim={1.5cm 19cm 4cm 5cm},clip]{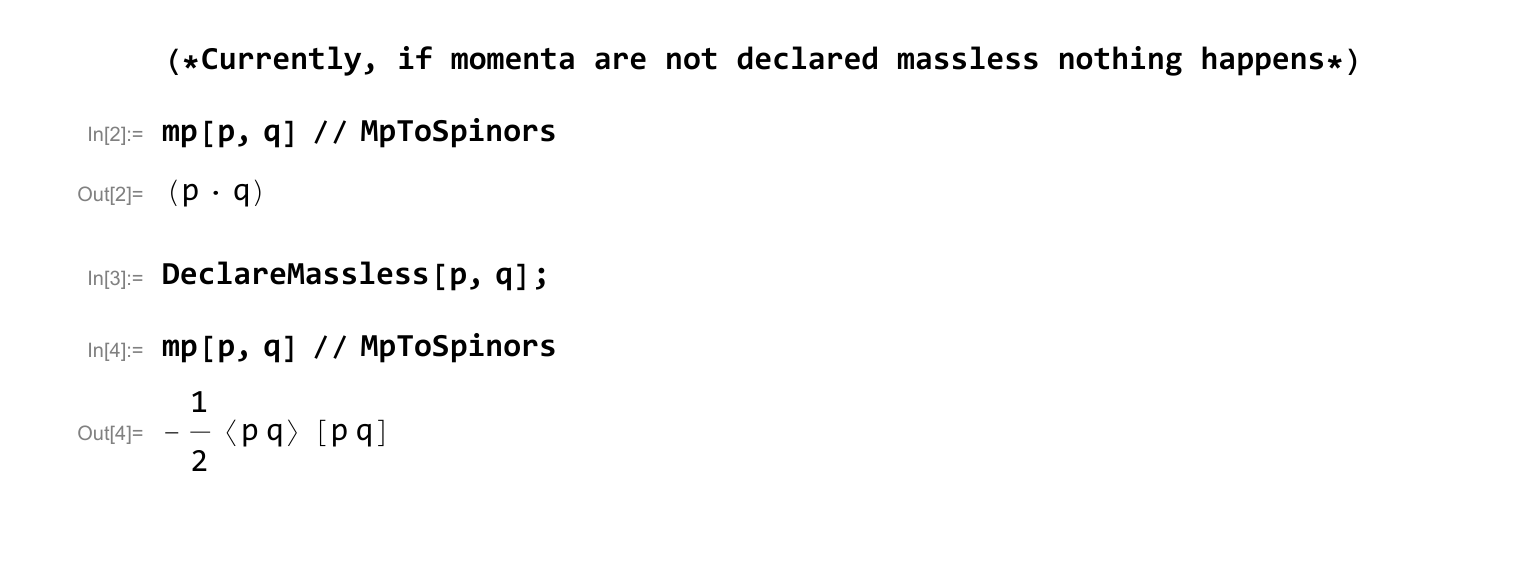}}}
	\end{center}
\end{figure}
Below, the function when applied to scalar products including polarization vectors.
\begin{figure}[H]
	\begin{center}
		\makebox[\textwidth]{
			\fbox{\includegraphics[page=1,trim={1.5cm 11.5cm 4cm 11.5cm},clip]{Example_MpToSpinors}}}
	\end{center}
\end{figure}

\subsection{CompleteDenominators}
The function \texttt{CompleteDenominators} completes individual squares brackets in denominators to full Mandelstam invariants.
The current implementation of this functions only works when we are dealing with a fully massless case, in fact if $p_i = |i\rangle[i| + \rho_i|\bar{i}\rangle[\bar{i}|$ and $p_j = |j\rangle[j| + \rho_j|\bar{j}\rangle[\bar{j}|$ then $\agl{i}{j}\sqr{j}{i} \neq s_{ij}$, and reconstructing the invariat from the products of brackets becomes far more involved.
\begin{figure}[H]
	\begin{center}
		\makebox[\textwidth]{
			\fbox{\includegraphics[page=1,trim={1.5cm 18.5cm 4cm 4.5cm},clip]{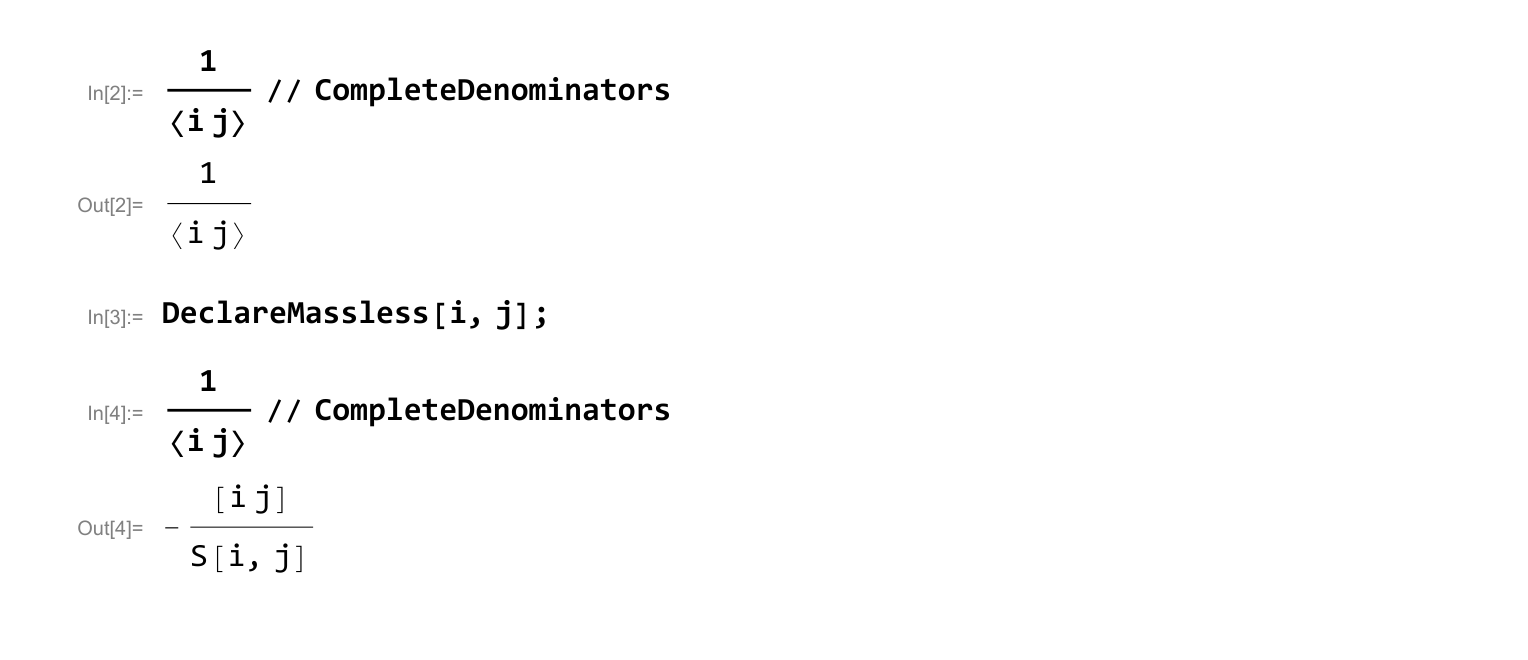}}}
	\end{center}
\end{figure}

\section{Numerics}\label{sec:numerics}

In most applications it is extremely convenient to be able to perform numeric checks of various nature, this might include for example numerically comparing two very different analytic expression or make sure that a result involving some reference spinor is independent of the chosen reference. To take a step further, numerical evaluations can be used to extrapolate analytic results through functional reconstruction. Then, cumbersome and complicated intermediate expressions are reduced to mere numbers, while focus is put only on the final quantity of interested. This approach is particularly powerful when combined with numeric evaluations over finite fields \cite{Peraro:2016wsq}, which provide exact numeric results (not subject to precision loss), while still maintaining good computational performances.

In this section we introduce the functions of the package which are related to numeric evaluations. The most important functions of the section are \texttt{GenSpinors}, which is used to generate numeric kinematics for a specified momentum configuration, and \texttt{ToNum} which converts a given analytic expression into the corresponding numerical value.

\subsection{ToNum}

The function \texttt{ToNum} translates a given analytic expression into the corresponding number, based on a numerical kinematic configuration generated through \texttt{GenSpinors}. The general logic is as follows:
\begin{itemize}
	\item One uses \texttt{GenSpinors} in order to generate some numeric kinematics (for details on 
	
	\texttt{GenSpinors} see Section \eqref{sec:genspinors}), which is then stored in the background. At this stage only two-component Weyl spinors are numerically computed.
	\item To each analytic object we associate a corresponding alias in which numeric values are stored. For example \texttt{SpinorUndot} $\mapsto$ \texttt{SpinorUndotN}, where the first is an analytic object while the second is vector of two numbers.
	\item By applying \texttt{ToNum} to an analytic expression, all the functions appearing in it are replaced by their numeric aliases, producing a numeric output. At this stage, the fundamental objects generated by \texttt{GenSpinor} (\textit{i.e.} the uncontracted spinors) are combined into derived quantities such as the spinor brackets. In order to optimise performances, these derived quantities are only computed if encountered in an expression, and then stored for later use.
	\item The stored numeric kinematics, both the fundamental as well as the derived objects, can be cleared through \texttt{ClearKinematics}
\end{itemize} 
Below is a simple example.
\begin{figure}[H]
	\begin{center}
		\makebox[\textwidth]{
			\fbox{\includegraphics[page=1,trim={1.5cm 19.5cm 4cm 4.5cm},clip]{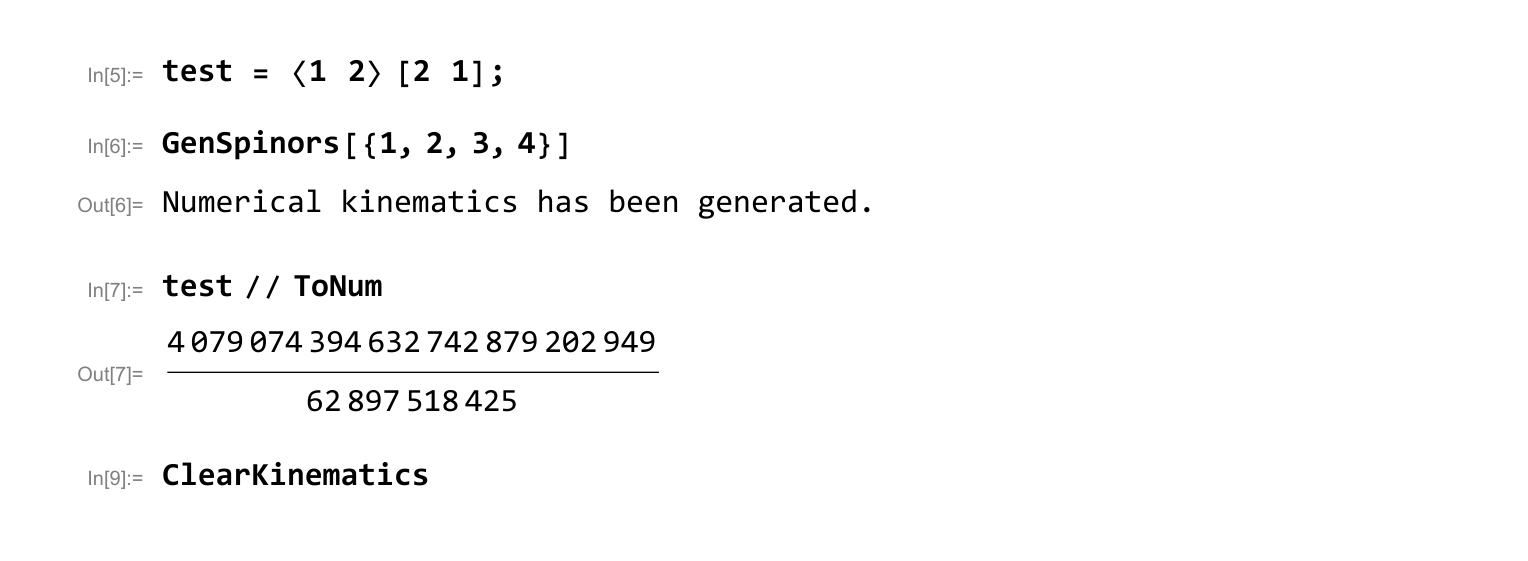}}}
	\end{center}
\end{figure}

\subsection{GenSpinors}\label{sec:genspinors}


Generation of numerical kinematics to us means finding a set of numbers to associate to all the spinors describing a given process, such that momenta are on-shell and conserved\footnote{We will follow a convention of all out-going momenta, thus conservation will always result in the momenta summing to zero.}. Since we use spinors as building blocks, the on-shell condition is automatically satisfied in the massless case and can easily be accounted for in the massive case. In order to satisfy momentum conservation one could make use of momentum-twistor variables \cite{Hodges:2009hk}: upon mapping the spinors to the twistor space both on-shellness and momentum conservation are trivialised thus any random set of numbers is suited to describe the kinematics (see for example \cite{Elvang:2013cua} for more details). However, since we want to accommodate several different scenarios including the massless and massive case, parametric kinematics, degenerate (three-point) and constrained kinematics, we found it easier to opt for a different approach. We work in spinor space, and upon randomly generating most of the components, we account for momentum conservation and other constraints by solving a system in the final variables, which have been carefully chosen so that they are always rational combinations of the previously fixed variables. This procedure allows a large degree of flexibility in the type of kinematics one can generate, while also retaining the benefit of rational number output.

\texttt{GenSpinors} only has one mandatory argument, being a list of momentum labels for which kinematics needs to be generated, the function however allows for a great number of options. In the following subsections we give a detailed description of all these options, along with their intended use and possible limitations. Before going into the details we stress once again that the generated kinematics is always complex.

\subsubsection*{AllMassless: Massive vs massless states}

When \texttt{GenSpinors} is provided only with the mandatory argument of a single list of labels, it is implicitly assumed that all the states should be massive and that no particular relation between the masses exists.
\begin{figure}[H]
	\begin{center}
		\makebox[\textwidth]{
			\fbox{\includegraphics[page=1,trim={1.5cm 16.5cm 4cm 4.5cm},clip]{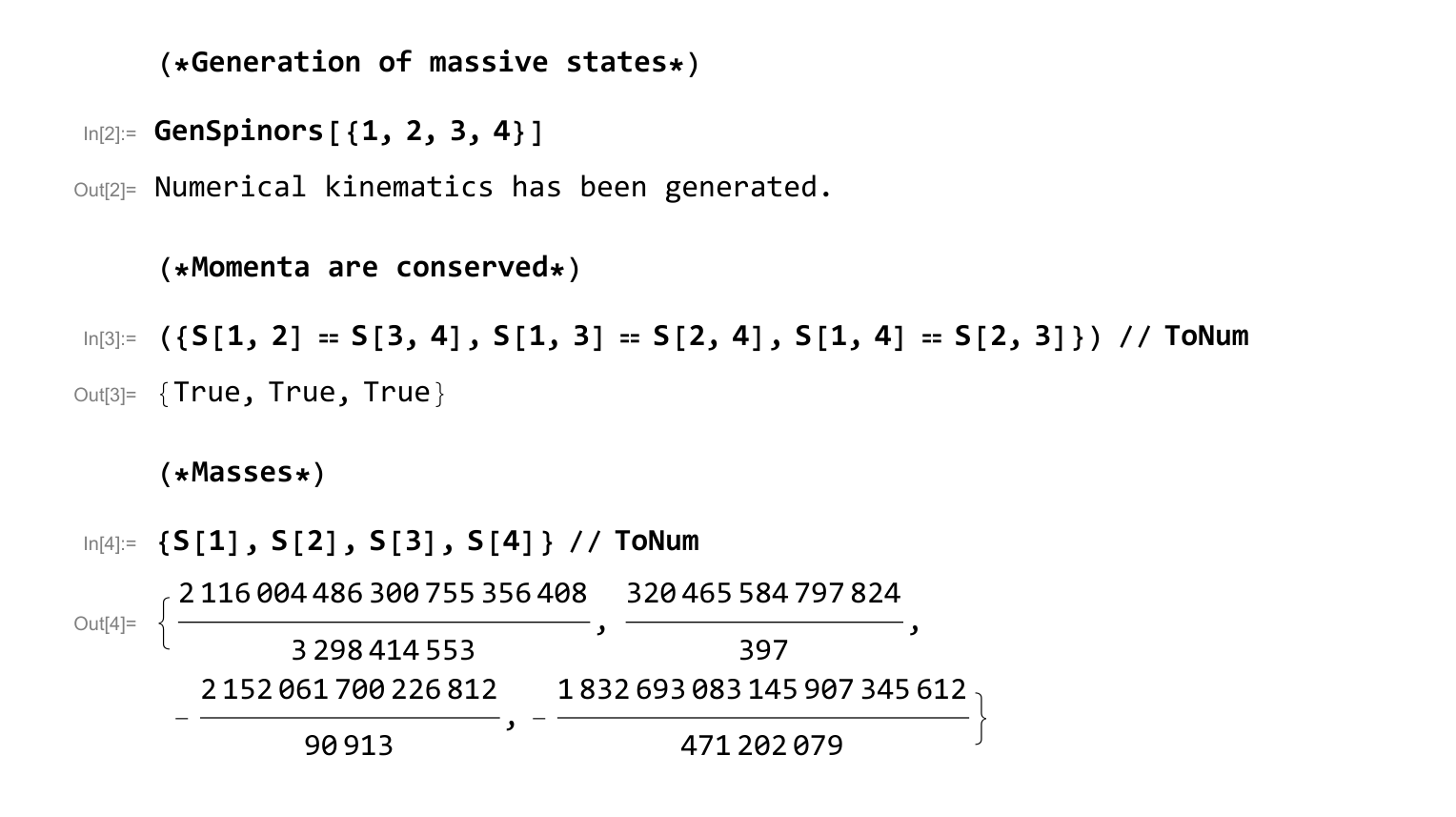}}}
	\end{center}
\end{figure}
Using the option \texttt{AllMassless} it is possible to generate fully massless kinematics, while by providing a list of two lists as inputs one can generate a mixed massive and massless process.
\begin{figure}[H]
	\begin{center}
		\makebox[\textwidth]{
			\fbox{\includegraphics[page=1,trim={1.5cm 8.5cm 4cm 14cm},clip]{Example_GenSpinors1}}}
	\end{center}
\end{figure}
Consecutive use of \texttt{GenSpinors} does not clear existing kinematics, it will only overwrite existing labels which reapper in a new call.
\begin{figure}[H]
	\begin{center}
		\makebox[\textwidth]{
			\fbox{\includegraphics[page=2,trim={1.5cm 6cm 4cm 17.5cm},clip]{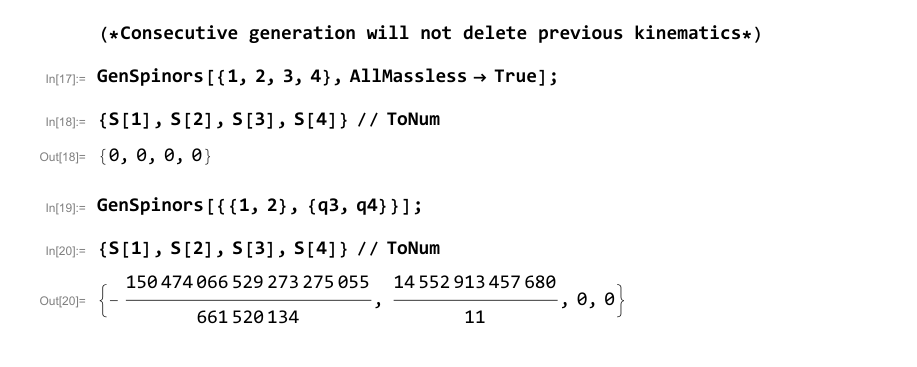}}}
	\end{center}
\end{figure}

\subsubsection*{SameMasses}
Using the option \texttt{SameMasses} it is possible to declare some of the masses to be the same, this option takes either a single list as input (if only one set of masses is required to be the same) or a list of lists allowing for different subsets of momenta having the same masses. 
\begin{figure}[H]
	\begin{center}
		\makebox[\textwidth]{
			\fbox{\includegraphics[page=2,trim={1.5cm 13.5cm 4cm 10cm},clip]{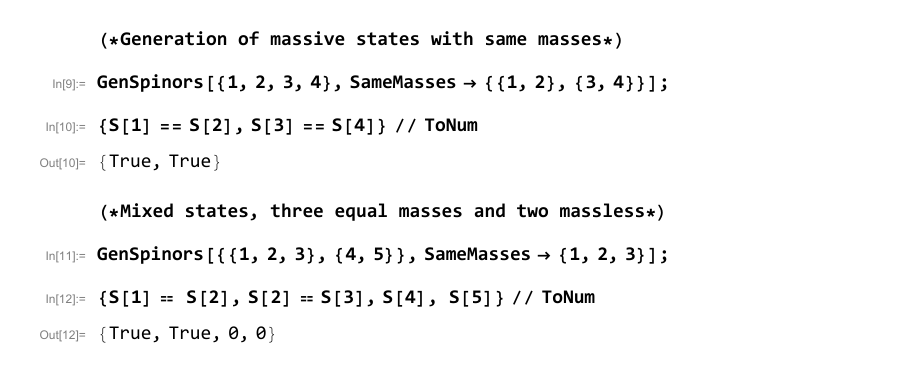}}}
	\end{center}
\end{figure}

\subsubsection*{DisplaySpinors}

The boolean option \texttt{DisplaySpinors} controls whether or not the numerical spinors are displayed by \texttt{GenSpinors} upon generation. The default value is False, and prompts the function to simply return a string stating the completion of the kinematics' generation.
\begin{figure}[H]
	\begin{center}
		\makebox[\textwidth]{
			\fbox{\includegraphics[page=1,trim={3cm 19cm 2.5cm 4.5cm},clip]{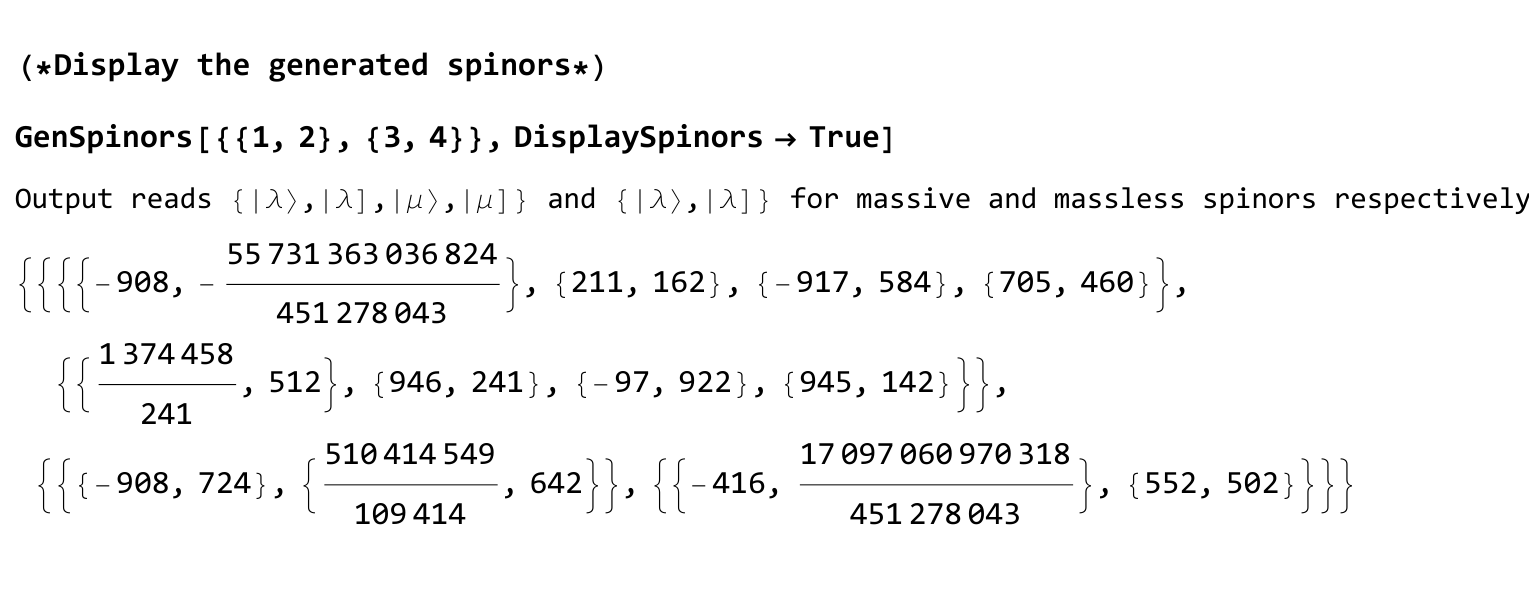}}}
	\end{center}
\end{figure}

\subsubsection*{Seed}
The numerical kinematics is generated through \texttt{Mathematica}'s random number generator. In order for the same numeric result to be repeatable it is possible to seed the random generator through the option \texttt{Seed}. Since this option will simply pass its value to \texttt{Mathematica}'s built in \texttt{SeedRandom}, it accepts the same values as the latter. Below, upon use of \texttt{Seed}, repeatedly generated kinematics leads to the same numbers.
\begin{figure}[H]
	\begin{center}
		\makebox[\textwidth]{
			\fbox{\includegraphics[page=1,trim={3cm 11cm 2.5cm 11cm},clip]{Example_GenSpinors2}}}
	\end{center}
\end{figure}

\subsubsection*{ParameterRange}

It is possible to set the interval within which random numbers are generated when feeding into \texttt{GenSpinors}. This is done through the option \texttt{ParameterRange}, whose default value is $10^3$, and which accepts either a single number $n_1$ (range will be 0 to $n_1$) or a list of two numbers $\{n_1,n_2\}$ (range will be between $n_1$ and $n_2$) as input. It is important to stress that this option only controls the range of the randomly generated parts of the kinematics, meaning that the spinor components which are fixed by solving momentum conservation conditions (or other constraints imposed on the kinematics through other options of \texttt{GenSpinors}) are not bound by this range.
\begin{figure}[H]
	\begin{center}
		\makebox[\textwidth]{
			\fbox{\includegraphics[page=1,trim={3cm 4.5cm 2.5cm 20cm},clip]{Example_GenSpinors2}}}
	\end{center}
\end{figure}

\subsubsection*{RationalKinematics}

For a variety of applications, including for example the already mentioned functional reconstruction on finite fields \cite{Peraro:2016wsq}, it is convenient to generate kinematics on the field of rational numbers. While this is the default for \texttt{GenSpinors}, it is possible to control whether kinematics is generated on $\mathbb{Q}$ or $\mathbb{R}$ through the option \texttt{RationalKinematics}.
\begin{figure}[H]
	\begin{center}
		\makebox[\textwidth]{
			\fbox{\includegraphics[page=1,trim={3cm 20cm 2.5cm 5cm},clip]{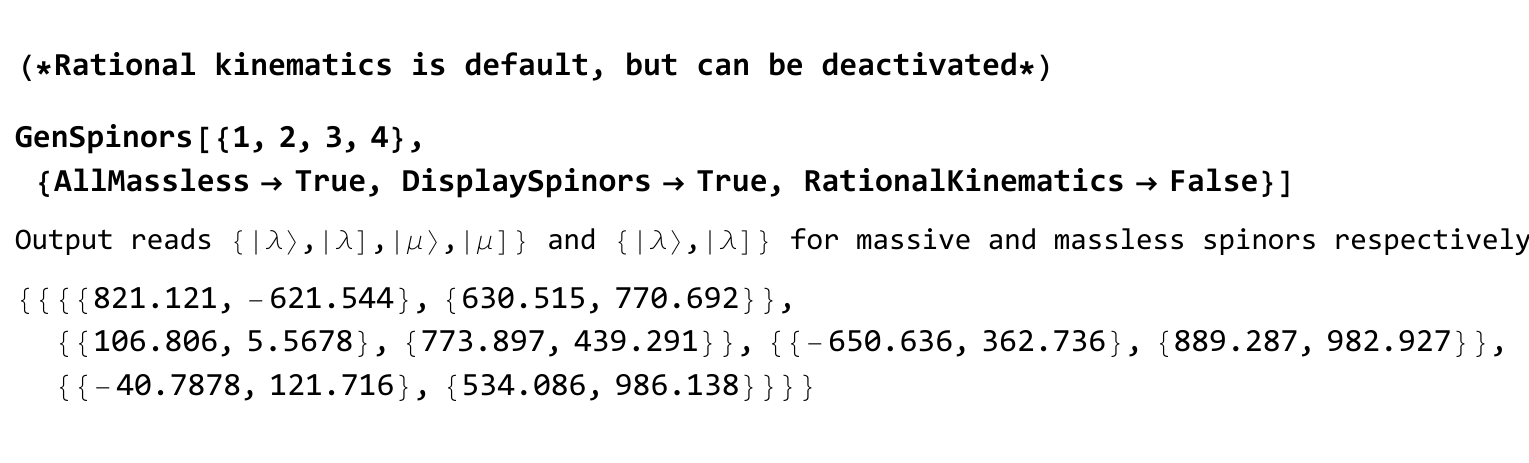}}}
	\end{center}
\end{figure}

\subsubsection*{Parametric and ParameterName}

Considering for example an $n-$particle massless process, it is well known that it can be described by a minimal set of $3n-10$ independent parameters. This can be accounted for in the numeric generation through the boolean option \texttt{Parametric}, whose default value is False. Setting it to True will generate a set of spinor components which are parametrized in terms of this minimal set of parameters called \texttt{\$par}.
\begin{figure}[H]
	\begin{center}
		\makebox[\textwidth]{
			\fbox{\includegraphics[page=1,trim={3cm 12.5cm 2.5cm 11.5cm},clip]{Example_GenSpinors3}}}
	\end{center}
\end{figure}
The name of the parameters can be change through the option \texttt{ParameterName}.
\begin{figure}[H]
	\begin{center}
		\makebox[\textwidth]{
			\fbox{\includegraphics[page=1,trim={3cm 4.5cm 2.5cm 18.5cm},clip]{Example_GenSpinors3}}}
	\end{center}
\end{figure}

\subsection*{Three-point kinematics}

The three-point kinematic is somewhat special due to the fact that, based on the helicity of the involved particles, two different types of kinematics are manifested. The two scenarios, both of them satisfying $p_+p_2+p_3=0$ and $s_{ij}=0$ and both of them being allowed only because we consider complex kinematics\footnote{Alternatively one could have considered a different signature of space time \cite{Witten:2003nn}.}, consist of $\agl{i}{j}=0$ and $\sqr{i}{j}\neq 0$ for every $i$, $j$, or the opposite. Which three-point kinematics is numerically generated is controlled by the option \texttt{Type3pt} which admits values \texttt{\$angle} and \texttt{\$square}.
\begin{figure}[H]
	\begin{center}
		\makebox[\textwidth]{
			\fbox{\includegraphics[page=1,trim={1.5cm 15.5cm 4cm 4.5cm},clip]{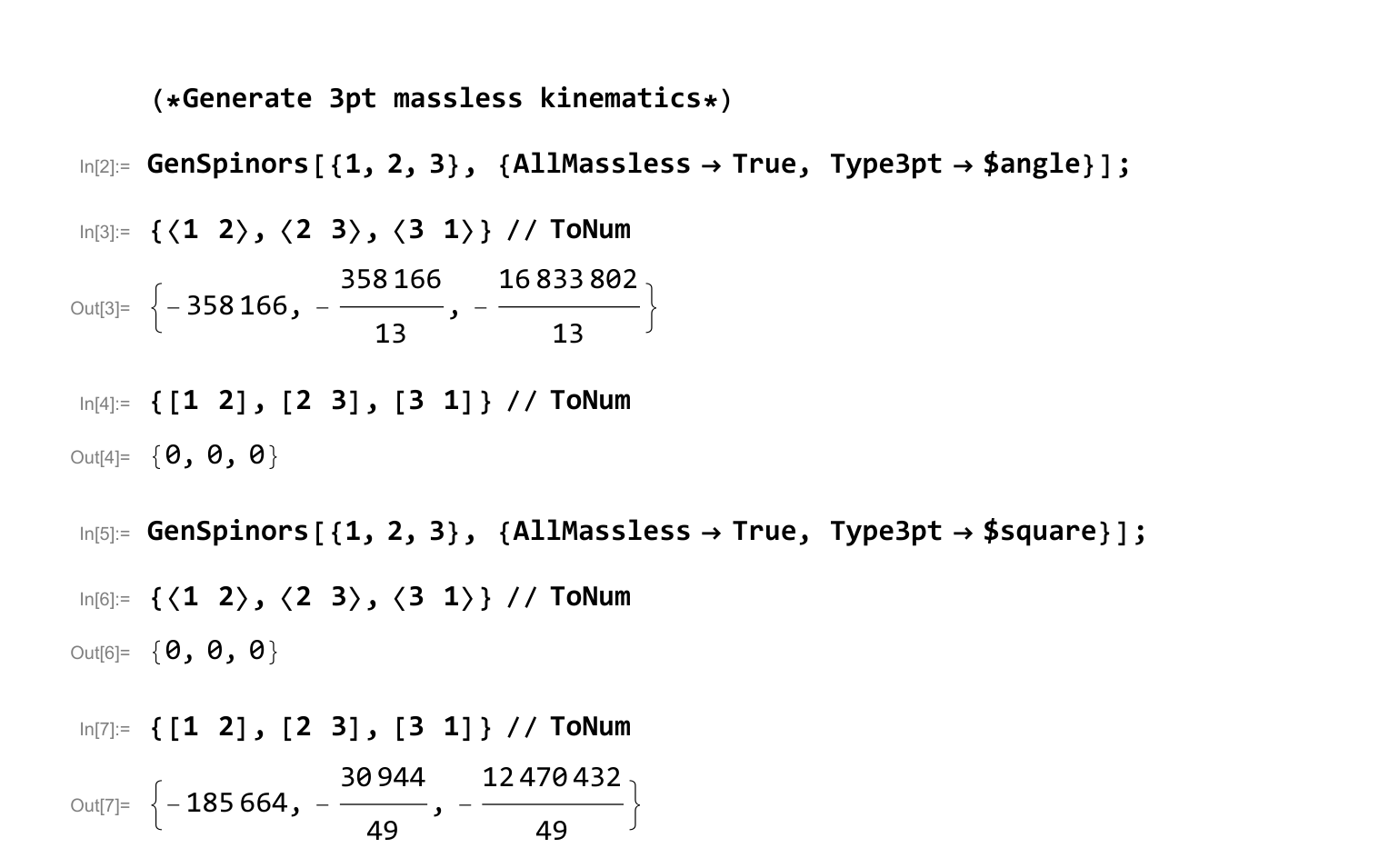}}}
	\end{center}
\end{figure}

\subsubsection*{SetMomentum}

In certain instances it might be convenient to set newly generated values of momenta to be equal (or opposite) to already generated momenta. This can be done through the option \texttt{SetMomentum}, which admits as a value a list of $m$ multiples of momentum labels, and fixes the first $m$ input labels through the given relation. An example is given below, where we consider a unitarity double cut of a four point amplitude.
\begin{figure}[H]
	\begin{center}
		\makebox[\textwidth]{
			\fbox{\includegraphics[page=1,trim={3cm 10cm 2.5cm 15.5cm},clip]{Example_GenSpinors4}}}
	\end{center}
\end{figure}
The given relationship between the momenta can be specified through any arbitrary number, but symbolic relations are not allowed and will prompt an error message.
\begin{figure}[H]
	\begin{center}
		\makebox[\textwidth]{
			\fbox{\includegraphics[page=1,trim={1.5cm 5.5cm 4cm 20.5cm},clip]{Example_GenSpinors4}}}
	\end{center}
\end{figure}

\subsection*{Acknowledgments}

We would like to thank Stefano De Angelis for his constant feedback on the early versions of the package, as well as for the collaboration on related topics, and Rodolfo Panerai for inspiring the original implementation of the code. We would also like to thank Gabriele Travaglini, Andreas Brandhuber and Sebastian P\"ogel for useful discussions and feedback, and Pierpaolo Mastrolia for discussions on \verb|S@M|. This project has received funding by the European Union's Horizon 2020 research and innovation programme under the Marie Sk\l{}odowska-Curie grant agreement No.~764850 {\it ``\href{https://sagex.org}{SAGEX}''}.

\newpage
\bibliographystyle{utphys}
\bibliography{biblio}

\end{document}